\documentclass[twocolumn, final,superscriptaddress]{revtex4-1} 
\usepackage{mathptmx}      
\usepackage{amsmath}
\usepackage{marvosym, pifont}
\usepackage{graphicx}
 \usepackage[colorlinks==true,
            pdftex,
             allcolors=blue,
             bookmarks=false,
             hypertexnames=true,
             linkcolor=blue]{hyperref} 
              
\usepackage{titlesec}
\usepackage{ctable}
\usepackage{subfigure}
\usepackage{adjustbox}
\usepackage{longtable}
\usepackage{multirow}
\usepackage{mathtools} 
\def\vecfc#1{\mathbf{#1}}
\def\fcb#1{{\color{black}{#1}}}
\def\etal{{\textit{et al.}}}

\def\VB{Veldhuis \& Biesheuvel}

\renewcommand\vecfc{\mathbf}

\begin{document}

\title{Path instabilities and drag in the settling of single spheres}


\author{Facundo Cabrera-Booman}\email{f.cabrera.booman@gmail.com}
\affiliation{Department of Mechanical and Materials Engineering, Portland State University, Portland, Oregon, USA.}
\affiliation{Univ Lyon, ENS de Lyon, CNRS, Laboratoire de Physique, F-69342 Lyon, France}
\author{Nicolas Plihon}
\affiliation{Univ Lyon, ENS de Lyon, CNRS, Laboratoire de Physique, F-69342 Lyon, France}
\author{Mickaël Bourgoin}
\affiliation{Univ Lyon, ENS de Lyon, CNRS, Laboratoire de Physique, F-69342 Lyon, France}

\date{April 2023}

\begin{abstract}
The settling behavior of individual spheres in a quiescent fluid was studied experimentally. The dynamics of the spheres was analyzed in the parameter space of particle-to-fluid density ratio ($\Gamma$) and Galileo number ($\mathrm{Ga}$), with $\Gamma \in (1.1, 7.9)$ and $\mathrm{Ga} \in (100, 340)$. The experimental results showed for the first time that the mean trajectory angle with the vertical exhibits a complex behavior as $\mathrm{Ga}$ and $\Gamma$ are varied. Numerically predicted regimes such as Vertical Periodic \fcb{for low $\Gamma$ values}, and Planar Rotating for high $\Gamma$ values were validated. In particular, for the denser spheres, a clear transition from planar to non-planar trajectories was observed, accompanied by the emergence of semi-helical trajectories corresponding to the Planar Rotating Regime. The spectra of trajectory oscillations were also quantified as a function of $\mathrm{Ga}$, confirming the existence of oblique oscillating regimes at both low and high frequencies. The amplitudes of the perpendicular velocities in these regimes were also quantified and compared with numerical simulations in the literature. The terminal velocity and drag of the spheres were found to depend on the particle-to-fluid density ratio, and correlations between the drag coefficient and particle Reynolds number ($Re_p$) as a function of Ga were established, allowing for the estimation of drag and settling velocity using $\mathrm{Ga}$, a control parameter, rather than the response parameter $Re_p$.
\end{abstract}

\maketitle

\section{Introduction}
Particles in fluids are representative of many natural and industrial systems and therefore extensively investigated in a variety of scenarios such as turbulence \cite{Cabrerainstpart, thesisfacu,mininni2020}, and low \cite{notre_slenderbodies,OBLIGADO2022103704} to moderate~\cite{zhoupaper,jdb2004} Reynolds number such as this work.
Particularly, and despite its apparent simplicity, the physics of finite size spheres settling hides a hierarchy of rich intricate phenomena, some of which are still shrouded in mystery. We are for instance still unable to finely model and predict the terminal velocity of a particle settling in a turbulent environment. The role of linear and non-linear drag~\cite{bib:good2014_JFM,Rosa2016}, the link with possible scenarios enhancing the settling~\cite{bib:maxey1987_JFM} or hindering it~\cite{bib:nielsen1993}, the influence of finite size effects~\cite{bib:chouippe2019} and the role of collective effects~\cite{bib:aliseda2002_JFM} are just some examples of subtle couplings which still need to be further explored to improve our capacity to predict the turbulent settling of spherical particles.
Challenges are particularly important for environmental issues such as the forecast of particle and pollutants deposition in the atmosphere, rivers and seas.

Interestingly, even the non-turbulent situation, where a sphere settles in a quiescent fluid, is already far from trivial and results in a series of path instabilities~\cite{jdb2004} not yet fully understood. These path instabilities are related to a complex wake dynamics which emerges for a sphere with a relative velocity with respect to the surrounding fluid. It is indeed well known for instance that the wake behind a fixed sphere of typical size $d$, in a steady stream with velocity $U$ and viscosity $\nu$, has a number of bifurcations that depend on Reynolds number $\mathrm{Re} = U d / \nu$. 
These transitions have been thoroughly explored in numerical and theoretical \cite{fabre1,tomboulides_orszag_2000,natarajan_acrivos_1993} and experimental \cite{nakamura, ormieresprovansal} studies for the case of fixed spheres in a steady stream for which the onsets of different wake bifurcations are finely characterised.

When the sphere is not fixed (e.g. if it is settling under gravity or rising due to buoyancy in a quiescent fluid), these wake instabilities develop into path instabilities~\cite{ern} as the momentum and torque exerted by the perturbed fluid onto the particle will influence its trajectory.
A pioneering work regarding fluidised beds already highlighted the non-applicability of Newton's free settling law on rising particles \cite{karamanev}, caused by the aforementioned wake effect on the particle trajectory. 
Jenny and coworkers~\cite{jdb2003,jdb2004} made the first systematic numerical study exploring the trajectory dynamics of a single spherical particle settling or rising in a quiescent unconfined fluid. This study was refined later by Zhou and Du\v{s}ek~\cite{zhoupaper}. The complex dynamics of rising or settling spheres has also been characterized experimentally and theoretically~\cite{pnas1,pnas2,auguste_magnaudet_2018,horowitz,veldhuis,breugem_new}.

Two dimensionless numbers control the free sphere settling problem: particle-to-fluid density ratio $\Gamma = \rho_p/\rho_f$ (with $\rho_p$ and $\rho_f$ the particle and fluid densities respectively) and Galileo number $\mathrm{Ga} = \sqrt{|\Gamma-1|g)}d_p^{3/2}/\nu$ (with $d_p$ the particle diameter, $g$ the local acceleration of gravity and $\nu$ the kinematic viscosity of the surrounding fluid). The Galileo number was defined here as $\mathrm{Ga} = U_{g}d_p/\nu$, where the characteristic velocity is the buoyancy velocity $U_{g}=\sqrt{|\Gamma-1|gd_p}$. The different regimes and bifurcations of single settling or rising spheres were then assessed in a $\Gamma$ -- $\mathrm{Ga}$ parameter space.
While the regimes observed for both density ratio below one (rising spheres and bubbles) \cite{pnas1,pnas2,karamanev,auguste_magnaudet_2018} 
and for density ratio above unity (particle settling) \cite{zhoupaper,horowitz,veldhuis,breugem_new,jdb2004}
are interesting, we will restrict ourselves to density ratios larger than unity in the present article. To keep this introduction concise, a detailed review of previous investigations is provided in Sec.~\ref{sec:results}, to which our experimental observations are systematically compared. We specifically stress that a number of important regions of the parameter space still remain experimentally unveiled and need to explored in order to characterise the settling regimes and corroborate numerical predictions. This is particularly the case for particle-to-fluid density ratios larger than 3.9 for which no experimental data is available.

Besides the complexity of path instabilities, the drag force experienced by the particles is an important element of the problem which has interested the scientific community. Inquiring in particular on whether the drag force of fixed spheres in a steady stream could be used to estimate the terminal settling or rising velocity of freely moving particles. Raaghav \etal~\cite{breugem_new} have studied the drag of rising and settling particles and concluded that for density ratios between 0.86 and 3.9, the particle settling drag estimated from the mean vertical terminal velocity of the spheres does not differ significantly from that of a fixed sphere in free stream flowing at the same velocity. The latter implies that the drag coefficient $C_D$ does not depend on particle-to-fluid density ratio. This idea is used extensively in the literature, and it has been widely used to obtain correlations and empirical models assuming a simple dependency of $C_D$ on particle Reynolds number $\mathrm{Re}_p = v_p d_p/\nu$~\cite{brown,dragcubes}. This has been proven incorrect for light particles where a marked dependency appears when $\Gamma < 0.1$~\cite{karamanev,auguste_magnaudet_2018}. \\
Another practical issue is that the correlations for drag and settling velocity available in the literature are usually given in terms of the particle Reynolds number. However, when the particles are free to move, the velocity $v_p$ is not a control parameter but a response parameter. For the case of settling particles, these correlations do not allow to give an explicit expression for the terminal velocity in terms of the drag coefficient $C_D$, because $C_D$ itself depends on the terminal velocity. However, from a pure dimensional analysis approach, the natural expected dependencies of the drag coefficient for settling spheres are both on $\Gamma$ and $Ga$, which are actual control parameters, only depending on known physical parameter of the problem (densities of the particles and the fluid, fluid viscosity, particle diameter and acceleration of gravity). This brings the two following questions: (i) to which extent is the approximation of $C_D$ not depending on density ratio valid? And (ii) can a correlation of $C_D$ be given in terms of $\mathrm{Ga}$ rather than $\mathrm{Re_p}$? This would allow to know the drag coefficient \textit{a priori} without requiring to know the terminal velocity beforehand.

In the present article, we investigate experimentally the settling of spherical particles in a quiescent fluid over a broad region of the parameter space, namely $1<\Gamma<8$ and $100<\mathrm{Ga}<350$ (symbols in figure~\ref{fig:parametersspace} indicate all points explored in the parameter space). For all the investigated conditions, we fully characterize the trajectory properties of the particles as well as the drag coefficient derived from the particle's terminal velocity.
The article is organized as follows. We first introduce the experimental setup in Sec.~\ref{sec:expsetup}. The results are then described in Sec.~\ref{sec:results}. Finally, our conclusions are summarized in Sec.~\ref{sec:conclusions}.
\section{Experimental Methods\label{sec:expsetup}}
\begin{figure}[!htbp]
\centering
 \includegraphics[width = 0.8\linewidth]{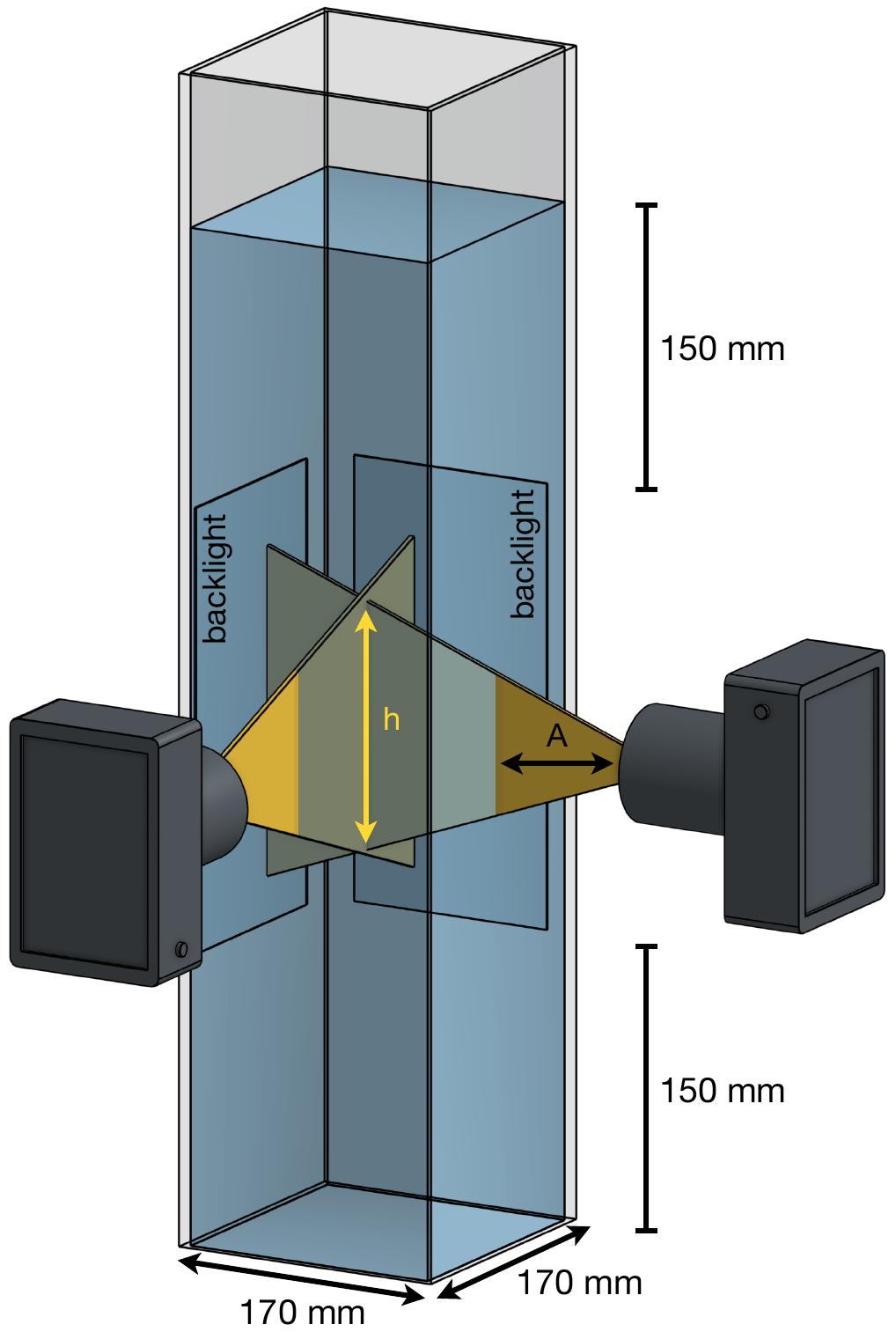}
 \caption{Experimental setup. Two cameras image the particles settling inside the water tank.}
\label{fig:expsetup_quiescent}
 \end{figure}
\subsection{Experimental setup and protocol}
The experiments are performed in a transparent PMMA tank with a square cross-section of $170\times170$~mm$^2$ and a height of $710$~mm, shown in Figure~\ref{fig:expsetup_quiescent}. The tank is filled with different mixtures of pure glycerol (Sigma-Aldrich W252506-25KG-K) and distilled water, ranging from 0\% to 40\% glycerol concentration. The viscosity of each mixture is measured with a rheometer Kinexus ultra+ from Malvern industries with a maximum uncertainty of \fcb{0.6\%}. 
The kinematic viscosity $\nu$ ranges from $10^{-6}$ to $1.05\times10^{-3}$~m$^2/$s. 
Moreover, as the viscosity is dependent on the temperature, an air-conditioning system keeps a constant room temperature of $(22\pm0.6)^{\circ}$C yielding a \fcb{2\%} uncertainty on the precise value of the viscosity. \\
A $150$~mm region of fluid above and below the visualisation volume is set to ensure both the disappearance of any initial condition imposed on the particles release and the effects of the bottom of the tank. Furthermore, a minimum distance of 20~mm between the tank walls and the particles is maintained. In this configuration and using the correlations proposed by Chhabra \etal~\cite{walleffect} the settling velocity hindering due to wall effects is estimated to be lower than 3\%.\\
The trajectory of the settling particles is recorded using two high speed cameras (model fps1000 from The Slow Motion Camera Company \textit{Ltd}) with a resolution of $720\times1280$~px$^2$ and a frame rate of 2300~fps. The movies recorded from these two cameras allow the implementation of time resolved 4D-Lagrangian Particle Tracking (4D-LPT) to reconstruct the particle trajectories~\cite{micaPTV}. 
\fcb{This method tracks particles with an uncertainty of $90\mu$m  which is estimated from the disparity between rays when stereo-matching the particle between the two cameras. This experimental noise on the particle position is short time correlated and gets significantly reduced by the high temporal redundancy associated to the oversampling achieved with the frame rate of 2300Hz and the subsequent gaussian filtering of the trajectories (further detailed below) used to estimate particle velocity. As a consequence, the uncertainty on the instantaneous velocity along trajectories is less than 4~mm/s \cite{micaPTV} while the associated uncertainty for the velocity averaged over a given trajectory drops below a few hundred microns per second.} 
Backlight illumination was used, with two LED panels facing each camera on the opposite side of the tank, as represented by the dark blue rectangles in Fig.~\ref{fig:expsetup_quiescent}.

Various series of experiments were carried with different optical magnification ratios, in order to access large scale properties of the trajectories (with lower magnification) as well as higher resolution data (with higher magnification). The magnification was varied by keeping the same optics mounted on the cameras, and varying the distance $A$ from the cameras to the exterior of the tank's wall. The datasets corresponding to these different situations are detailed in the next subsection. 

 \begin{figure*}[ht!]
 \centering
 \includegraphics[width=\textwidth]{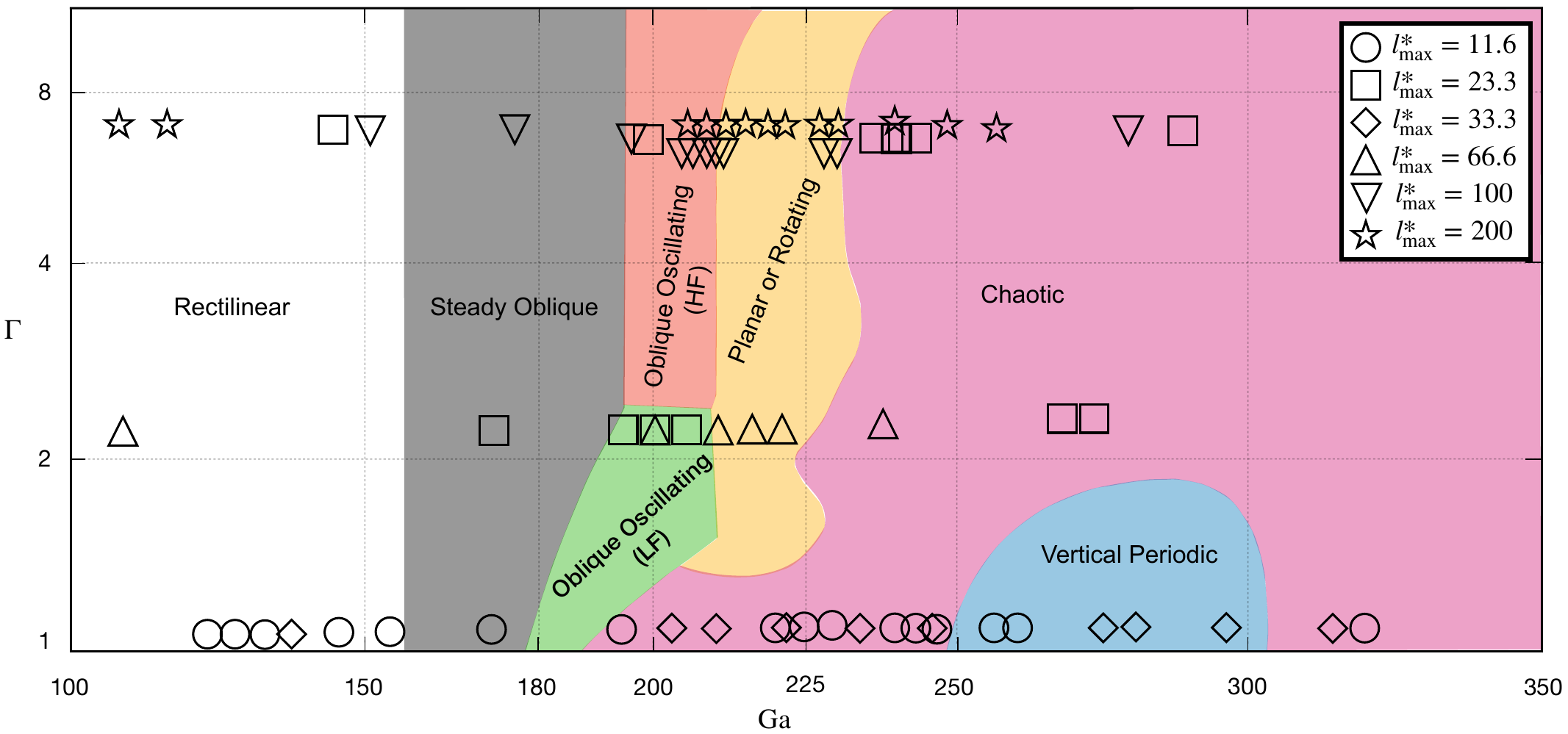}
 \caption{Particle-to-fluid density ratio ($\Gamma$) -- Galileo number ($\mathrm{Ga}$) space of parameters. Data points are classified by their maximum trajectory length $l^\ast_\mathrm{max} = h/d_p$.} 
\label{fig:parametersspace}
 \end{figure*}
In order to span the $\Gamma$ - $\mathrm{Ga}$ parameters space, we considered a set of spherical particles with different diameters ($d_p$) and densities ($\rho_p$), while varying the water-glycerol mixture in order to vary the fluid viscosity $\nu$. Varying the fluid viscosity $\nu$ allows to change $\mathrm{Ga}$ for a given type of particle, at the expense of the slight modification of the value of $\Gamma$ due to the associated variation of the fluid density. The characteristics of the particles and the ranges of values $\mathrm{Ga}$ and $\Gamma$ investigated in this articles are reported in Table~\ref{table:particlesQF}.
Overall, a total of 68 points in the $\Gamma$ - $\mathrm{Ga}$ parameters space has been explored (see figure~\ref{fig:parametersspace}. For each point up to 25 independent drops were released in order to test the repeatability of the observed regimes and the eventual presence of bi-stable regions where different settling regimes could co-exist in the same region of the parameters space. 
\begin{table}[ht]
\centering
\resizebox{0.49\textwidth}{!}{
\begin{tabular}{c c c  c c c c}
\specialrule{0.15em}{0em}{0em} 
\it{Material (label)}  & $\rho_p$~(kg/m$^3$) & $d_p$ (mm) & $\Gamma$ & $\mathrm{Ga}$ & Ra($\mu$m) \\ \specialrule{.15em}{0em}{0.1em} \vspace{0.5mm} 
 Metal &7950 & $\{1, 2, 3\}$ & 6.6-7.8 &  112-290 & 9 \\  \vspace{0.5mm} 
 Glass  &2500 & $3$ &  2.1-2.5 & 130-270 & 15 \\  
 \fcb{Polyamide}  &1150 & $6$ &1.1-1.3 & 124-340 & 120  \\ \specialrule{.05em}{0em}{0.5em} 
\end{tabular}}
\caption{Properties of the different settling particles investigated. See text for details.}
\label{table:particlesQF}
\end{table}
The particle's diameter and sphericity were measured using a microscope with a precision of $10~\mu$m. In particular, no significant deviation from the spherical shape or the manufacturer's documented diameter could be measured. The surface roughness of the particles was also measured, with a Scanning Electron Microscope ZEISS SUPRA 55 VP, over an area of $200\times500$~$\mu$m$^2$. The arithmetical mean height of rugosities $\mathrm{Ra}$ reported in table~\ref{table:particlesQF} shows a high degree of smoothness as $\mathrm{Ra}/d_p < 0.05$, therefore roughness is not expected to alter the spheres dynamics~\cite{surfaceroughness2}. \fcb{In particular, the following particles were used: Metal - \textit{Stainless Steel Ball AISI 316 Grade 100} from COMAC Europe; Glass - \textit{Soda Lime Grade 60} and Polyamide - \textit{PA 6.6 Grade 2} both from Marteau \& Lemari\'e.}

The experimental procedure is the following: the tank is filled with a water-glycerol mixture and after approximately 24 hours the temperature at different positions in the fluid's bulk differs in less than \fcb{$0.6^{\circ}C$} thus thermal equilibrium is reached. Then a standard calibration of the 4D-LPT system is performed \cite{micaPTV}. The spheres are released at the center of the tank with \fcb{standard Stainless Steel Anti-acid and Anti-magnetic} chemical tweezers. The tweezers are completely submerged below the air-liquid interface and released after approximately 20~s when the fluid free surface is at rest. A minimum time of $120$~s is taken between successive drops to ensure that the fluid has no perturbations left from the previous drop. \fcb{The waiting time is chosen to be at least 12 viscous relaxation times $\tau = d_p^2/\nu$. Note that the viscous times vary between different cases and the resulting waiting time is in between 12$\tau$ and 1000$\tau$, with a median value of 150$\tau$.} 
\subsection{Data sets}
%
The experiments were conducted using two different optical magnifications, resulting in various values of the non-dimensional trajectory length $l^\ast_\mathrm{max} = h/d_p$ ranging from 11.6 to 200 (see Fig.~\ref{fig:expsetup_quiescent}).
Note that the lowest values of $l^\ast_\mathrm{max}$ (11.6 and 23.3) correspond to the larger optical magnification, or small $A$ (hence giving better spatial resolution, but shorter tracks) while the larger values of $l^\ast_\mathrm{max}$ were obtained with the smaller magnification, or large $A$ (resulting in a larger field of view, hence giving access to longer trajectories, what is important in particular to properly estimate the frequency of oscillating regimes). The values of $l^\ast_\mathrm{max}$ are reported in the $\Gamma - \mathrm{Ga}$ parameters space in Fig.~\ref{fig:parametersspace}. 

All the relevant geometric (inclination and planarity) and dynamic characteristics (spectral content and terminal velocity) of particle trajectories cannot be equally addressed from the different datasets as the accuracy of their estimate depends on the maximum accessible track length $l^\ast_\mathrm{max}$. Empirically, we found that to reasonably resolve trajectory inclination, a dimensionless trajectory length of at least $l^\ast\gtrsim 10$ (which is accessible with all datasets) is needed. This has been tested by checking the estimation of the inclination angle using the longest trajectories in the oblique regime and successively considering shorter and shorter portions of those long tracks.
On the other hand, the quantification of the planarity via the eigenvalue method detailed in Sec.~\ref{sec:results}, requires $l^\ast\gtrsim 23$ - a condition not met for plastic particles, due to their large diameter.
This conclusion has been reached by checking the estimation of the planarity using the longest available trajectories in the chaotic regime and successively considering shorter and shorter portions of those long tracks. This effect will be explored further in Section~\ref{sec:planarity}. 
Finally, the spectral analysis required long trajectories, an issue further discussed in Sec.~\ref{sec:spheresoscillations}. 

In order to reduce experimental noise (due to inevitable particle detection errors in the Lagrangian Particle Tracking treatment~\cite{bib:ouellette2005_ExpFluids}), the raw trajectories are smoothed by convolution with a Gaussian kernel of width $\sigma = 12$~frames. It behaves as a low-pass filter with a cut-off frequency $f_c = \rm{fps}/\sigma= 2300~\rm{Hz}/\sigma = 192$~Hz.
Spectral analysis is therefore expected to be well resolved for frequencies up to of the order of 80~Hz as to respect the Nyquist-Shannon sampling theorem. 

As previously mentioned, for each data point in the $\Gamma$ - $\mathrm{Ga}$ parameters space, at least 10 and up to 25 experimental repetitions were executed and their trajectories analysed. This is mandatory in order to test the repeatability of the observed regimes, estimate uncertainties, and eventually detect multi-stable regions of the parameters space where multiple settling regimes may coexist. The uncertainties in quantities extracted from this data (e.g. trajectory angle or planarity) are taken as the standard deviation over the total set of drops for each data point. \fcb{For computed quantities (i.e. Reynolds number, Galileo number and Drag coefficient) the errors are estimated from a standard propagation of errors, see for instance \cite{breugem_new}. }\\

Finally, in the remainder of this article, dimensionless parameters are denoted by a superscript asterisk. Spatial variables are normalized by particle diameter $x^{\ast} = x / d_p$, velocities are normalized by the buoyancy velocity $v^{\ast} = v / U_g  = v/\sqrt{|\Gamma-1| g d_p}$, and time is normalized by the response time of the particles $\tau_g = d_p/U_g$.

\section{Results\label{sec:results}}
In this section, we first recall and present the  different settling regimes reported in the literature. Then, the features of the 68 points experimentally investigated in the parameter space (Fig.~\ref{fig:parametersspace}) are described. Particular emphasis is put on their geometric and spectral properties, as well their terminal velocity and drag coefficient estimation

\subsection{Different Regimes}

The different regimes in the parameters space obtained from numerical simulations by Zhou and Du\v{s}ek~\cite{zhoupaper} are represented by different colors in Fig.~\ref{fig:parametersspace}. Seven distinct regimes were numerically identified, whose features are summarized in the following:  
\begin{enumerate}
\item Rectilinear Regime (white), with planar vertical trajectories and no inclination or oscillations;
\item Steady Oblique Regime (gray), with planar and oblique trajectories with respect to the vertical, and no oscillations; 
\item Oblique Oscillating Regime, with planar and oblique trajectories, and the presence of oscillations. The frequency of oscillations $f^\ast$ depends on the particle-fluid density ratio $\Gamma$, with a High-Frequency Regime (HF, orange) at $f^{\ast}\simeq0.18$ and a Low-Frequency (LF, green) at $f^{\ast}\simeq0.068$.
\item Planar or Rotating Regime (yellow), a \fcb{bi}-stable region of the parameters space composed of oblique and (High or Low-Frequency) oscillating  trajectories, which could be either planar or exhibit a slowly rotating symmetry plane (thus generating helicoid-like trajectories), coexisting with Chaotic Regimes. The High-Frequency Regime, Low-Frequency Regime, and Chaotic Regime coexist in this zone.
\item Vertical Periodic Regime (blue), where the trajectories are planar, rectilinear and vertical, and oscillate at \fcb{ $f^{\ast}\in (0.141, 0.15)$};
\item and finally the Chaotic Regime (pink), with oblique and non-planar trajectories with no periodic oscillations.
\end{enumerate}

\begin{figure*}[!tbp]
\centering
\includegraphics[width=\textwidth]{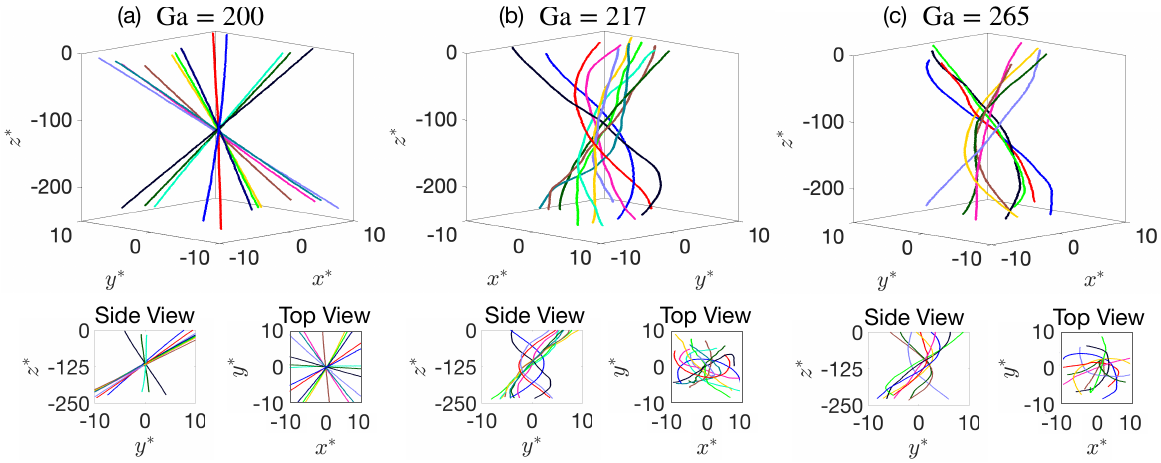}
 \caption{Typical trajectories \fcb{for particles with $\Gamma \approx 7.9$} in the: (a) Steady Oblique; (b) Planar or Rotating; and (c) Chaotic Regimes.}
\label{fig:spheres_traj}
 \end{figure*}
A systematic study of the bifurcations between regimes was performed numerically by~\cite{zhoupaper}. That study narrowed down the limits between regimes, in terms of $\mathrm{Ga}$ and $\Gamma$, and has reported new regimes not previously detected in the simulations by Jenny \etal~\cite{jdb2003,jdb2004} (such as a Helical/Rotating Regime and a Vertical Periodic Regime). They also demonstrate the existence of bi-stable zone in the parameters space, where two regimes could co-exist. For instance, for moderate particle-to-fluid density ratios $\Gamma\lesssim 2$ a bi-stable regime between a Chaotic and a Vertical Oscillating Regime are reported, while for larger density ratios they report bi-stability between Planar Oscillating and Helical Regimes. Furthermore, they have better quantified trajectory parameters such as angle, velocities and spectral content. Note that this description of the dynamics of individual particles was later used as a benchmark for numerical investigations of collective particle effects~\cite{uhlmann, picano}.
Few analytical results have been derived regarding the bifurcations between different settling regimes, one exception being the transition between the Rectilinear and the Steady Oblique Regimes which have been analytically shown by Fabre \etal~\cite{fabre2} to occur at a critical Galileo number of the order of 155, independently of the particle-to-fluid density ratio, in excellent agreement with the numerical findings previously mentioned.

To the best of our knowledge, only three experimental studies \cite{veldhuis, horowitz,breugem_new} have explored the predictions made by aforementioned simulations and theories.
Horowitz \etal~\cite{horowitz} were mostly interested in regimes for rising spheres or slightly denser than the fluid and high Galileo numbers: they studied particle-to-fluid density ratios $\Gamma$ below 1.4 and Galileo numbers ranging from $10^2$ to $10^4$. In particular, they studied trajectory angle and drag following the work of Karamanev~\cite{karamanev}. Intriguingly, most findings from this study deviate from numerical simulations by Zhou and Du\v{s}ek~\cite{zhoupaper}, in particular for the case of settling particles which will be investigated here. On the other hand, Veldhuis and Biesheuvel~\cite{veldhuis}, although with some discrepancies, observed several of the dynamical regimes observed in the numerical simulations. In particular, oblique trajectories with no significant frequencies (Steady Oblique Regime in simulations) were reported. They also report oblique trajectories with oscillations at three dominant dimensionless frequencies of 0.07, 0.017 and 0.025 (Oblique Oscillating Regime in simulations), whose presence depends on the particle-to-fluid density ratio $\Gamma$. Finally, an oblique chaotic regime with no dominant frequencies and random trajectory curvature (Chaotic Regime) was described. These regimes were measured for particle-to-fluid density ratios $\Gamma$ of 1.3 and 2.3 at various Galileo numbers spanned by varying the fluid viscosity. Finally, in 2022, Raaghav \etal~\cite{breugem_new} performed experiments on rising and settling particles, with four particle-to-fluid density ratios ($\Gamma$ = 0.87, 1.12, 3.19 and 3.9) and $\mathrm{Ga}$ ranging from 100 to 700. They confirmed and contradicted some results of previous numerical simulations and experiments. The low $\mathrm{Ga}$ regimes (up to the Steady Oblique Regime) is unambiguously confirmed, in agreement with previous studies. For higher Galileo numbers (typically above 200), they found however discrepancies both with previous numerical and experimental studies. For instance, they observed a bi-stable behavior (between the Oscillating and Chaotic Regimes) for moderately dense spheres ($\Gamma \simeq 1.1$) in the range $250<\mathrm{Ga}<300$ in agreement with by Zhou and Du\v{s}ek~\cite{zhoupaper}, but for density ratios above 3, they did not observe the High-Frequency Oblique Oscillating Regime reported by Zhou and Du\v{s}ek~\cite{zhoupaper}; they confirmed though the existence of a helical mode, although no bi-stability with the Chaotic Regime was observed, contrary to the findings by Zhou and Du\v{s}ek~\cite{zhoupaper} reported.

Our experiments confirm the existence of all the predicted regimes, in regions of the parameters space in relatively good agreement with the ones delimited by numerical simulations. Figures~\ref{fig:spheres_traj}(a-c) qualitatively show some examples of trajectories.
More specifically, Fig.~\ref{fig:spheres_traj} (a-c) show some representative 3D trajectories for $\Gamma \approx 7.9$ particles, from the $l^\ast_\mathrm{max}= 200$ dataset. The trajectories have been arbitrarily centered in the horizontal axis. Sub-figures show top and side views. 

Fig.~\ref{fig:spheres_traj}(a) represents a case of planar and oblique type of trajectories measured here at $\mathrm{Ga}$ = 200. Note that steady and oscillating regimes are almost indistinguishable in such a representation by a simple visual inspection of the trajectories as the amplitude of oscillations is of the order of the particle diameter. The distinction between the two regimes will be quantitatively discussed later, based on the estimation of the particle velocity and their spectral analysis (the example shown in Fig.~\ref{fig:spheres_traj}(a) is actually an oblique oscillating case). It can also be noted that the angle of the trajectories with the vertical in this oblique regime remains almost constant for all drops (the angle will be quantitatively investigated in the next subsection, and is of the order of 5$^\circ$ in the present example), but each trajectory has its own direction so that the ensemble forms a cone hence preserving the global symmetry of the problem. 

Fig.~\ref{fig:spheres_traj}(b) represents a sample of trajectories of $\Gamma \approx 7.9$ particles at $\mathrm{Ga} = 217$. By combining the side and top views, it can be seen that several of these trajectories are consistent with portions of helicoids (for instance the red and the dark blue curves, which appears as quasi circular from the top view, although even with the $l^\ast_\mathrm{max}= 200$ dataset, we only catch half of the period at most).
Those co-exist with non-planar chaotic trajectories (as for instance the black and yellow curves). These measurements fall in the tri-stable regime previously mentioned. 

Finally, Fig.~\ref{fig:spheres_traj}(c) presents several trajectories that fall in the Chaotic Regime: all trajectories are different and no pattern of planarity or oscillations is present. 

After this brief qualitative description of some observed trajectory regimes, the next Subsections present a systematic quantitative analysis of the different properties used to characterise trajectory geometry and dynamics: angle with the vertical, planarity, spectral content, terminal velocity, and drag.
\subsection{Trajectories Angle}\label{sec:angle} 
\begin{figure}[!htbp]
 \centering
 \includegraphics[width = 0.87\columnwidth]{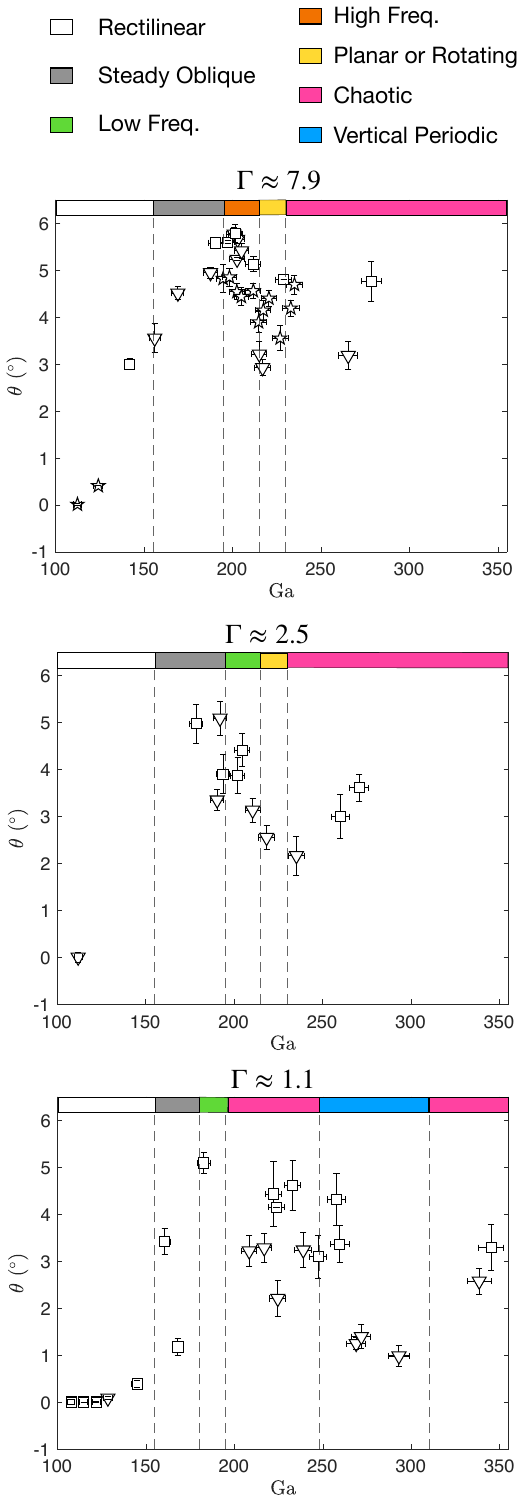}
\caption{Trajectory angle versus Galileo number for the three particle densities. Regimes are delimited by dashed vertical lines and identified by colors  following Fig.~\ref{fig:parametersspace}. Symbols represent the value of $l^\ast_\mathrm{max}$, according to Fig.~\ref{fig:parametersspace}.}
\label{fig:angle}
\end{figure}

For each recorded trajectory we define the settling orientation as the angle between a 3D linear fit of the trajectory and the vertical, and for each given set of parameters ($\mathrm{Ga},\Gamma$) we define the mean settling orientation as the ensemble average of settling angles over all trajectories recorded at those parameters. Fig.~\ref{fig:angle} shows the mean settling orientation as a function of $\mathrm{Ga}$ for the three different classes of particles investigated ($\Gamma \approx 7.9$, $\Gamma \approx 2.5$ and $\Gamma \approx 1.1$). Besides, the different settling regimes as reported from numerical simulations and previously shown in Fig.~\ref{fig:parametersspace} are delimited by the dashed vertical lines and identified by coloured rectangles that respect the colour code in Fig.~\ref{fig:parametersspace}. Furthermore, the type of symbols represents the value $l^\ast_\mathrm{max}$, also following the nomenclature of Fig.~\ref{fig:parametersspace}.

A smooth transition from rectilinear to oblique (primary regular bifurcation) is seen around the expected critical Galileo number of 150 for $\Gamma \approx 1.1$ and $\Gamma \approx 7.9$ particles and, although there is a lack of data points in this region of $\mathrm{Ga}$ for $\Gamma \approx 2.5$ particles, the available data points are consistent with a similar transition also occurring in the same range of $\mathrm{Ga}$ for those particles. More precisely, if the threshold between this regimes is defined as the Galileo number value at which the angle of the mean settling orientation has a non-zero angle, $\Gamma \approx 1.1$ and $\Gamma \approx 7.9$ particles present threshold values of $(125\pm10)$ and $(115\pm10)$ respectively, leading to a joint threshold at $\mathrm{Ga} = (120\pm15)$.
The trajectory angle is then found to continuously vary with the Galileo number; see for example $\Gamma \approx 7.9$ particles: the angle varies monotonously from 0 to 6 degrees in the $\mathrm{Ga}$ range 110-190. \fcb{With this respect, the transition between the rectilinear regime and the steady oblique regime in our experiment somewhat appears as an imperfect bifurcation rather than a sharp bifurcation with a critical Galileo number $Ga\approx 155$. The origin of such an imperfect bifurcation remains unclear and would deserve further future investigations.}

Additionally, the maximum observed angles are $(5.7^{\circ}\pm0.2^{\circ})$, $(5.1^{\circ}\pm0.2^{\circ})$ and $(5.1^{\circ}\pm0.2^{\circ})$ for density ratios 7.9, 2.5 and 1.1, respectively. This maximum angle is reached around $\mathrm{Ga} = 200$ in all cases in the region of parameters space that has been identified in numerical simulations by Zhou and Du\v{s}ek~\cite{zhoupaper} and previous experiments~\cite{horowitz, breugem_new} as corresponding to the Oblique Regimes, although the distinction between steady and oscillating regimes requires further analysis of the spectral content of the trajectories, which will be presented later. We note also that, although the detailed trend of the settling angle with $\mathrm{Ga}$ as presented here has not been systematically explored in previous studies, the values we observe for the maximum settling angle are in good agreement with the range of angles previously reported: ``of about 4 to 6 degrees" in the Steady Oblique and Oblique Oscillating Regimes in numerical simulations by Zhou and Du\v{s}ek~\cite{zhoupaper}, `` approximately $ 4^{\circ}$ to $7.5^{\circ}$ " in \cite{horowitz} and `` approximately $ 2.8^{\circ}$ to $7.4^{\circ}$ " in \cite{breugem_new}. 

It can be seen in Fig.~\ref{fig:angle} that for large Galileo numbers (typically $\mathrm{Ga}>200$) multiple values of the average settling angle can be observed for similar values of $\mathrm{Ga}$. These situations are generally consistent with regions of the parameters space which have been identified in numerical simulations either as multi-stable (yellow) or chaotic (pink). For the denser particles, such multi-values of the settling angle are for instance pronounced in the range $\mathrm{Ga}\in (200,230)$ encompassing both the HF-Oblique Oscillating (orange) and tri-stable Planar/Rotating (yellow) regions of the numerical parameters space, what may suggest that the multi-stable Planar/Rotating Regime, identified numerically around $\mathrm{Ga}\approx 220$, may actually extend further into the HF-Oblique Oscillating region at lower Galileo numbers. For the lightest particles, the trend to observe multiple values of the settling angle is very clear in regions of $\mathrm{Ga}$ expected to correspond to the Chaotic Regime (pink), in particular in the range $\mathrm{Ga}\in(200,260)$. For the intermediate density case ($\Gamma \approx 2.5$), this trend is observed in the vicinity of the LF-Oblique Oscillating Regime (green), what may be a sign that as for the dense particles case, the region numerically identified as bi-stable Planar/Rotating (yellow) may actually extend to lower values of $\mathrm{Ga}$ particularly into the LF-Oblique Oscillating region.

It is also interesting to see that for the $\Gamma \approx 1.1$ particles the drop of the settling angle in the range $\mathrm{Ga}\in(250,300)$ is consistent with the numerical prediction of a Vertical Periodic Regime (blue) appearing in that range and surrounded by Chaotic Regimes.

Overall, measured settling angles are consistent with what is expected from the numerical parameters space. With the exception of a probably more extended multi-stable region (yellow) overlapping (partially or totally) the Oblique-Oscillating regions. 

\subsection{Trajectories Planarity}\label{sec:planarity} 
\begin{figure}[!htbp]
\centering
   \includegraphics[width = 0.87\columnwidth]{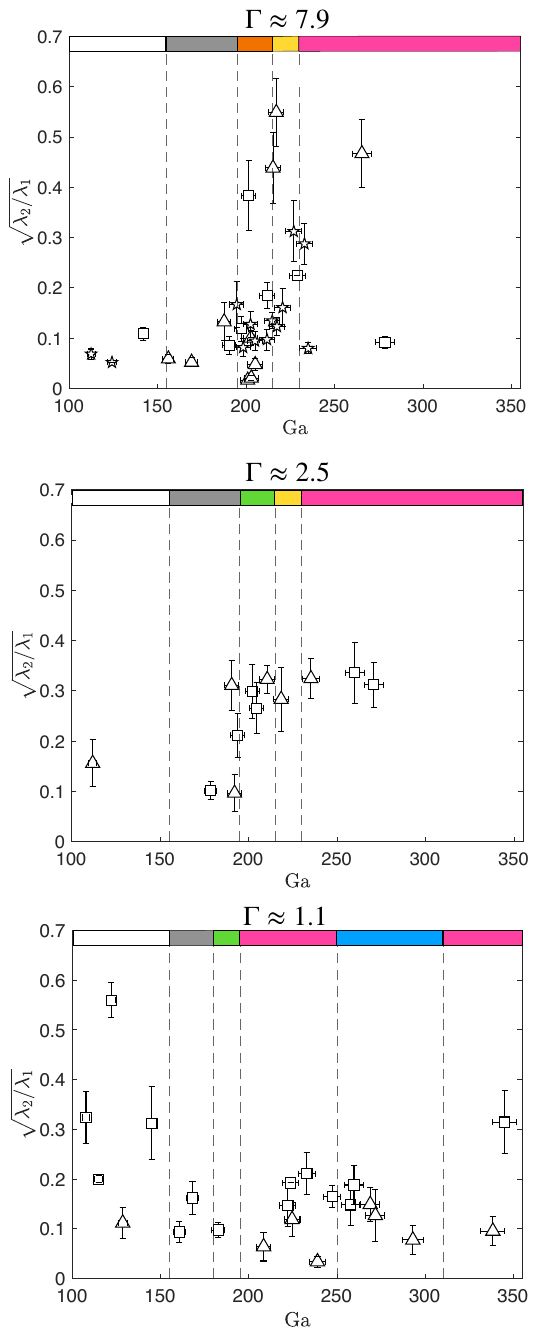}
\caption{Planarity versus Galileo number for the three particle densities. Regimes are delimited by dashed vertical lines and identified by colors following Fig.~\ref{fig:parametersspace}. Symbols represent the value of $l^\ast_\mathrm{max}$, according to Fig.~\ref{fig:parametersspace}.}
   \label{fig:planarity_met}
	\end{figure}

The trajectory planarity is quantified by the ratio of eigenvalues $\lambda_2/\lambda_1$ (with $\lambda_1 \geq \lambda_2$) of the dimensionless perpendicular (to gravity) velocity correlation matrix defined as:
\begin{equation}
\langle\vecfc{v}_{\perp}^{\ast}~\vecfc{v}_{\perp}^{\ast\rm{T}}\rangle=
\begin{bmatrix}
<{v_x^{\ast}}^2> & <v_x^{\ast} v_y^{\ast}>\\
<v_y^{\ast} v_x^{\ast}> & <{v_y^{\ast}}^2>
\end{bmatrix}
,
\end{equation}
with $v^{\ast} = v/U_g$. Perfectly planar trajectories yield $\lambda_2/\lambda_1$=0, while non-vanishing values of this ratio indicate a departure from planarity \cite{zhouthese}. Note that the analysis of the planarity only yields meaningful results for trajectories with  $l^\ast_\mathrm{max} > 33.3$. 
Fig.~\ref{fig:planarity_met} shows the ratio $\sqrt{\lambda_2/\lambda_1}$ versus $\mathrm{Ga}$ number, for the three types of particles.  
As in previous figures, the different regimes are delimited by dashed vertical lines and identified by coloured rectangles.

Planarity is lost at $\mathrm{Ga}$ = $(220 \pm 15)$ for $\Gamma \approx 7.9$ particles and at $\mathrm{Ga}$ = $(220 \pm 15)$ for $\Gamma \approx 2.5$ particles. At these points the ratio between the eigenvectors of the velocity correlation matrix $\sqrt{\lambda_2/\lambda_1}$ increases from approximately 0.15 to 0.50  for $\Gamma \approx 7.9$ particles (0.30 for $\Gamma \approx 2.5$ particles). 
The range of Galileo number where planarity is found to be lost is consistent with the transition towards the Planar or Rotating Regime reported in numerical simulations by Zhou and Du\v{s}ek~\cite{zhoupaper}, with a possible overlap with the LF-Oblique Oscillating region for $\Gamma \approx 1.1$ particles and with the HF-Oblique Oscillating region for $\Gamma \approx 2.5$ particles. On the other hand,  no clear transition between planar and non-planar trajectories is observed for $\Gamma \approx 1.1$ particles, which may be due to too small values of $l^{\ast}_\mathrm{max}$.

In the case of $\Gamma \approx 7.9$ particles, the loss of planarity seems to be associated to the emergence of helicoidal trajectories. Fig.~\ref{fig:spheres_traj}(b) presents indeed a sample of trajectories for $\Gamma \approx 7.9$ particles, representative of the ensemble of trajectories at $\mathrm{Ga}\sim 217$, that are consistent with a half-helicoid. Similar trajectories are found at $\mathrm{Ga}$ = \{215, 217, 221\} and $\mathrm{Ga}$ = \{228,233\}, for several values of $l^\ast_\mathrm{max}>33.3$. Hence the aforementioned loss of planarity for data with Galileo numbers larger than $(220\pm15)$ (see Fig.~\ref{fig:planarity_met}) can be related to the appearance of these helicoid-like trajectories. 
Limitations of the measurement volume, even in the $l^\ast_\mathrm{max} =200$ configuration, do not allow to be fully conclusive as only a portion of the helicoid's period is recognizable. However, assuming that these trajectories are helicoids, the radius of their horizontal projection (Fig.~\ref{fig:spheres_traj}(b) top view) would be roughly 7 particle diameters, and their pitch would be approximately 500~particle diameters. 
Similar helicoid-like trajectories have also been seen experimentally in previous studies, although for smaller density ratios, $\Gamma < 3.9$ (recall that metallic particles in the present study have a density ratio $\Gamma\simeq 7.5$, which has not been investigated in previous works): \cite{veldhuis} reported what are possibly helicoidal trajectories for particles with density ratio of the order of $\Gamma \simeq 2.5$ (hence close to the present $\Gamma \approx 2.5$ particles), while \cite{breugem_new} found similar trajectories for particles with $\Gamma = \{3.2,~3.9\}$. Their results show a pitch of the order of 430~$d_p$ which is comparable to the one of 500~$d_p$ found here.  
In this sense, the results of this work confirm the existence of such non-planar, very likely helicoidal, regime for $\mathrm{Ga}\in (215, 233)$ at larger particle-to-fluid density ratios, in the range of metallic particles ($\Gamma \approx 7.9$). Recall that the short $l^{\ast}$ in the data sets of $\Gamma \approx 2.5$ and $\Gamma \approx 7.9$ particles do not allow to see a portion of an helicoid long enough to make such claims.

Fig.~\ref{fig:planarity_met} for $\Gamma \approx 2.5$ an $\Gamma \approx 7.9$ particles also shows signatures of non-planarity in the region numerically identified as chaotic ($\mathrm{Ga}\gtrsim230$), in agreement with the sample trajectories shown in Fig.~\ref{fig:spheres_traj}(c), where several trajectories show a clear departure from simple portions of helicoids. A clear distinction between non-planar helicoidal and chaotic trajectories, with a systematic characterization of the pitch and radius of the helicoids and of the frontier with the Chaotic Regime would nevertheless require further dedicated experiments with a taller visualisation volume.
\subsection{Trajectories Oscillations}\label{sec:spheresoscillations} 
We analyze the emergence of oscillatory dynamics by studying the fluctuations of the horizontal (\textit{i.e.} perpendicular to gravity) dimensionless velocity: ${v'}_{\perp}^{\ast} := {v}_{\perp}^{\ast} - \langle{v}_{\perp}^{\ast}\rangle$.
In particular, while oblique-oscillatory regimes have been experimentally reported for density ratios below 3.9, we want to confirm here their existence at higher density ratios (\textit{i.e.} for the $\Gamma \approx 7.9$ particles, with $\Gamma = 7.9$) and in that case evaluate the corresponding frequency. On the other hand, the existence of a Vertical Periodic Regime (light blue region in Fig.~\ref{fig:parametersspace}) for density ratios below 1.8, as predicted by Zhou and Du\v{s}ek~\cite{zhoupaper} \fcb{was only very recently corroborated experimentally \cite{breugem_new}}. This regime is expected to have trajectories with zero angle and Low-Frequency Oscillations. Recall that the regime has been already discussed in the previous section where a sharp decrease in trajectory angle was found. We will therefore confirm here that the oscillations are at the Low-Frequency $f^{\ast} \approx 0.06$.

Numerical simulations by Zhou and Du\v{s}ek~\cite{zhoupaper} predict the existence of Oblique-Oscillatory Regimes for $\mathrm{Ga}$ of the order of 200, with a characteristic dimensionless frequency $f^\ast$ which depends on the density ratio $\Gamma$. More specifically, the simulations by Zhou and Du\v{s}ek~\cite{zhoupaper} predict a transition from a Low-Frequency Regime (with a dominant dimensionless frequency $f^{\ast} \approx 0.07$, corresponding to green regimes in previous graphs) to a High-Frequency Regime (with $f^{\ast} \approx 0.18$, corresponding to orange regimes in previous graphs) occurring at $\Gamma \approx 2.3$. However, previous experiments by Veldhuis and Biesheuvel ~\cite{veldhuis}~and Raaghav \textit{et al.} have only partially confirmed this scenario. Veldhuis and Biesheuvel~\cite{veldhuis}~for instance did observe Oblique-Oscillating Regimes in the expected range of Galileo number for particles with density ratios $\Gamma \approx 1.5$ and $\Gamma \approx 2.5$, but they report a dominant characteristic frequency of $f^\ast \approx 0.25$ for the lower density ratio case (\textit{i.e.} about three times higher than the numerical prediction) while two main frequencies, of the order of 0.07 and 0.25, were detected for the larger density ratio. On the other hand, Raaghav \textit{et al.} consistently report a Low-Frequency Oblique-Oscillating Regime (with $f^\ast\approx 0.06$) for particles with density ratio $\Gamma \approx 1.1$, but did not find any planar High-Frequency Oblique-Oscillating Regime for particles with $\Gamma = 3.9$, for which only non-planar helical trajectories (similar to those reported in the previous section of this work) were observed. The existence of Oblique-Oscillating Regimes (and eventually the value of their frequency) for high density ratios therefore remains open.

\begin{figure}[!htbp]
\centering
   \includegraphics[width = 0.5\textwidth]{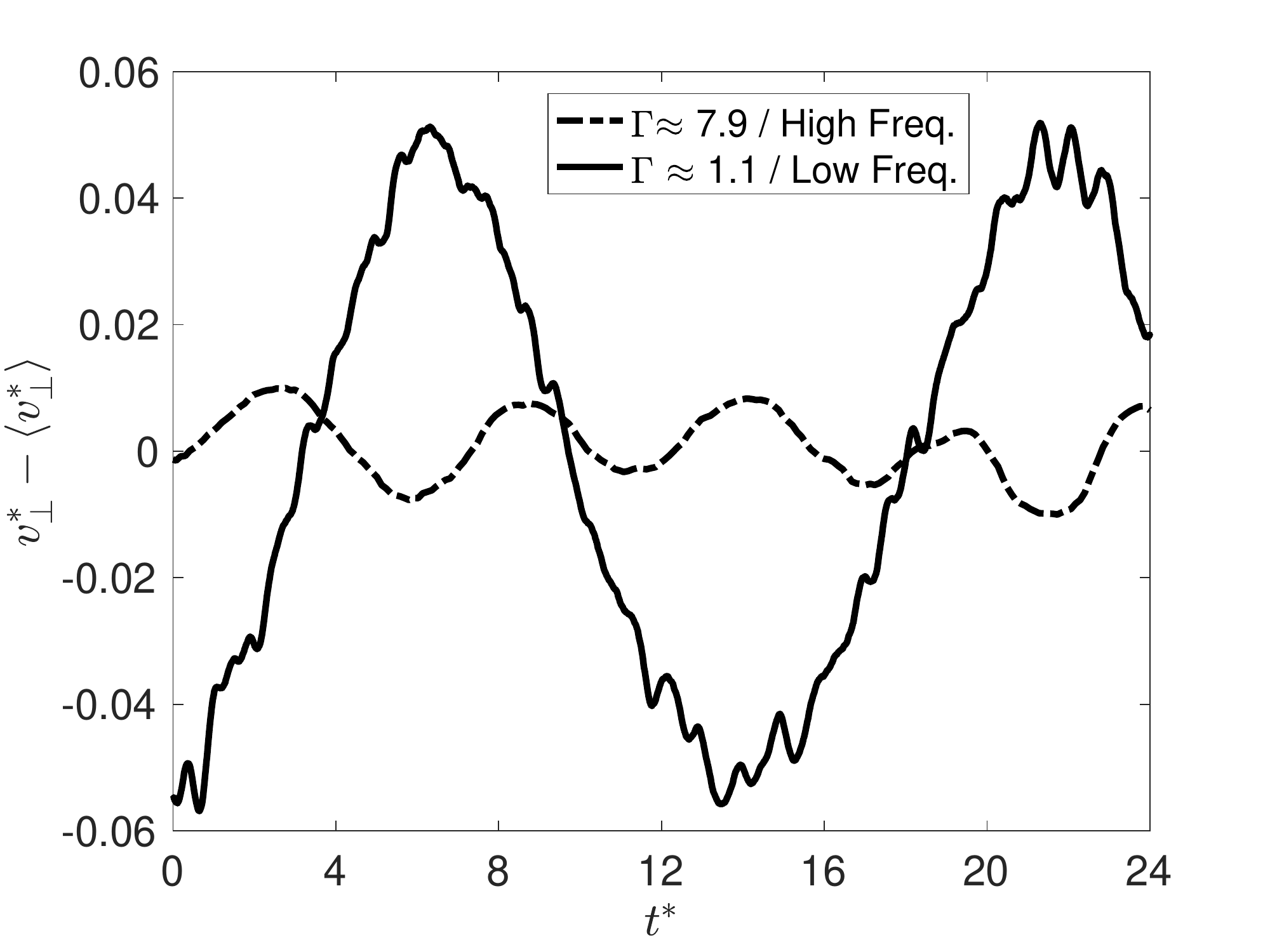}
\caption{Typical perpendicular (to gravity) velocity fluctuations for particles \fcb{with $\Gamma \approx 1.1$ and $\mathrm{Ga}=208$ that correspond to the Low-Frequency Regime (continuous line), and particles with $\Gamma \approx 7.9$ and $\mathrm{Ga}=200$ in the High-Frequency Regime (dashed line).}} 
  \label{fig:vperp_highvslow}
\end{figure}
\begin{figure}[!htbp]
\centering
 \includegraphics[width=1\linewidth]{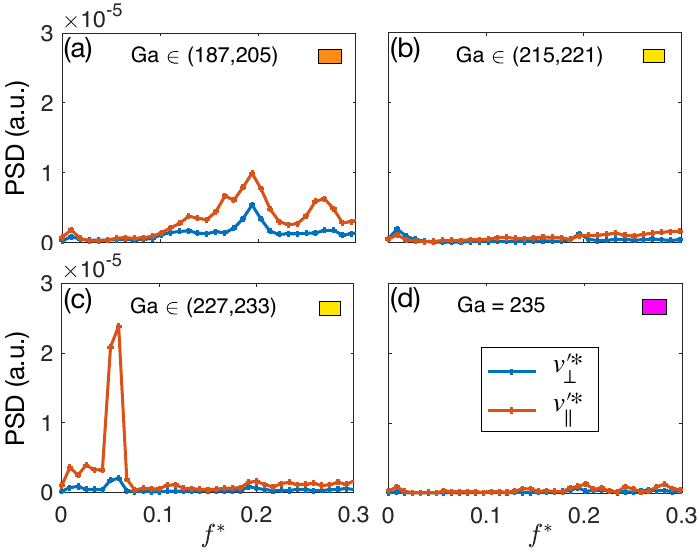}
\caption{PSD of parallel and perpendicular dimensionless velocity fluctuations (${v'}^{\ast}_{\parallel}$ and ${v'}^{\ast}_{\perp}$, respectively) for $\Gamma \approx 7.9$ particles. Colors correspond to Regimes defined in Fig.~\ref{fig:parametersspace}.}
\label{fig:fft}
\end{figure}
\begin{figure}[!htbp]
\centering
 \includegraphics[width=1\linewidth]{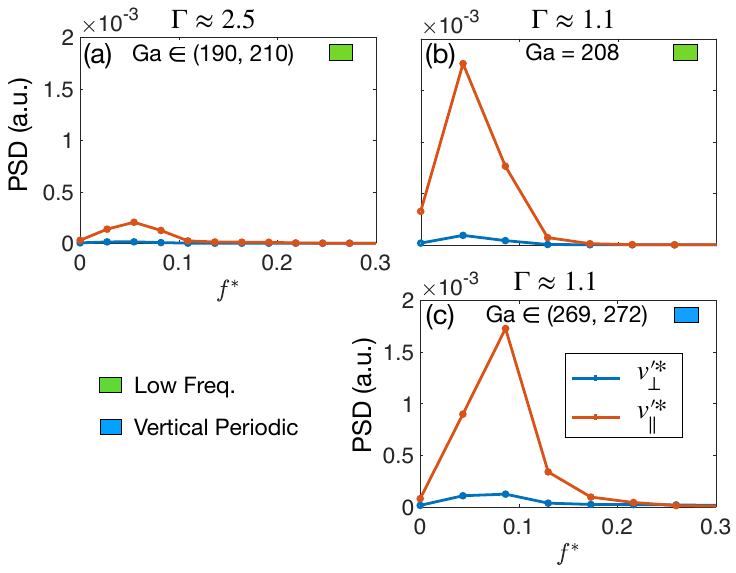}
\caption{PSD of parallel and perpendicular dimensionless velocity fluctuations (${v'}^{\ast}_{\parallel}$ and ${v'}^{\ast}_{\perp}$, respectively) for $\Gamma \approx 2.5$ particles in the LF-Oblique Oscillating Regime (a) and Vertical Periodic Regime (b) and (c) for $\Gamma \approx 1.1$ particles in the LF-Oblique Oscillating Regime.}
\label{fig:fft_PandG}
\end{figure}

Fig.~\ref{fig:vperp_highvslow}, shows a sample of perpendicular velocity fluctuations versus time for $\Gamma \approx 7.9$ particles  at $\mathrm{Ga}=200$, and $\Gamma \approx 1.1$ particles for $\mathrm{Ga}=208$. They exhibit a clear oscillatory dynamics, which is oblique (remember that $\theta \approx 5^{\circ}$ for particles at these $\mathrm{Ga}$) with marked frequency and amplitude differences. $\Gamma \approx 7.9$ particles show higher frequency and smaller amplitude than $\Gamma \approx 1.1$ particles. These observations are in qualitative agreement with numerical predictions.
The amplitude ratio between the High and Low-Frequency perpendicular dimensionless velocity oscillations of approximately 5 times is found however to be substantially smaller than what is reported in numerical simulations by Zhou and Du\v{s}ek~\cite{zhoupaper} where a ratio of 12 is observed. 
From the oscillations reported in Fig.~\ref{fig:vperp_highvslow}, it is possible to estimate the typical dimensionless frequencies $f^\ast$ for both regimes which is found to be of the order of 0.07 for the Low-Frequency case ($\Gamma \approx 1.1$ particles) and of the order of 0.2 for the High-Frequency case ($\Gamma \approx 7.9$ particles). These values are in good agreement with the numerical prediction, and the spectral analysis that follows.

A more accurate and systematic analysis of the oscillatory dynamics in the different regimes can be performed by computing the Power Spectral Density (PSD) of the velocity fluctuations averaged over multiple realizations in a narrow range of $\mathrm{Ga}$. 
Fig.~\ref{fig:fft} presents various PSD of velocity fluctuations at different values of the Galileo number, for the $l^\ast_\mathrm{max}= 200$ data-set of $\Gamma \approx 7.9$ particles. Both parallel and perpendicular components of velocity fluctuations have been analyzed. Each sub-figure presents the ensemble average of all PSDs in ranges of $\mathrm{Ga}$ where the spectral content was found to be robust:
$\mathrm{Ga}$ = \{187, 195, 198, 202, 205\} for Fig.~\ref{fig:fft}(a), $\mathrm{Ga}$ = \{215, 217, 221\} for Fig.~\ref{fig:fft}(b), $\mathrm{Ga}$ = \{227, 233\} for Fig.~\ref{fig:fft}(c), and $\mathrm{Ga}$ = 235 for Fig.~\ref{fig:fft}(d). We note that the spectral resolution, limited by the accessible trajectory length, is $0.01$. All measurements with $\mathrm{Ga}$ smaller than 187 have no spectral content (settling is then stationary, either vertical or oblique), therefore not shown.

The perpendicular velocity fluctuations PSDs presented in Fig.~\ref{fig:fft}(a) show that for $\mathrm{Ga}\in (187, 205)$ oscillations have a broad frequency peak centred around a dominant frequency $f^{\ast} = (0.19\pm0.01)$, and a secondary frequency around $f^{\ast} = (0.27\pm0.01)$.
The dominant frequency confirms the High-Frequency nature of the oscillations qualitatively discussed in the previous paragraphs for $\Gamma \approx 7.9$ particles at $\mathrm{Ga} = 200$, corresponding to the perpendicular velocity signal shown in Fig.~\ref{fig:vperp_highvslow}. 
It is also in agreement with the frequency predicted in numerical simulations by Zhou and Du\v{s}ek~\cite{zhoupaper} for such dense particles in this range of Galileo number, where a High-Frequency Oblique Oscillating Regime, with $f^{\ast} = 0.18$ has been reported by Zhou and Du\v{s}ek~\cite{zhoupaper}.

The main difference between these experiments and the simulations by Zhou and Du\v{s}ek~\cite{zhoupaper} is the non-negligible intensity of the peak at $f^{\ast} \approx 0.27$ (and possibly a sub-harmonic of $f^{\ast} \approx 0.13$). 
The existence of the frequency peak at $f^\ast = 0.27$ reminds of the observation by Veldhuis and Biesheuvel~\cite{veldhuis} who reported a similar frequency for particles both the Low and High-Frequency Regimes and was interpreted as a possible fourth harmonic of the Low-Frequency $f^\ast = 0.07$.

When $\mathrm{Ga}$ is increased to the range $(215,221)$, the trajectories lose any significant spectral signature. Neither the parallel, nor the perpendicular velocity PSD in Fig.~\ref{fig:fft}(b) show any marked peak. Only a mild peak at f$^\ast$ = $(0.01 \pm 0.01)$ is present for both parallel and perpendicular velocities and a mild peak at f$^\ast$ = $(0.19 \pm 0.01)$, with an intensity 6 times smaller than in the previous $\mathrm{Ga}$ range for the parallel velocity.
The angular and planarity analysis in the previous Subsection suggest that trajectories of $\Gamma \approx 7.9$ particles in this range of $\mathrm{Ga}$ might fall in the Planar or Rotating Regime, with some evidence of the existence on helicoidal trajectories in this regime. The estimated pitch of the helicoids ($\approx 500d_p$) would correspond to a frequency of oscillation of $f^{\ast} = v_{\parallel}^{\ast}/(500) \approx 0.002$, in principle out of reach of the 0.01 resolution of the present spectral analysis. The mild peak at $f^{\ast} \approx 0.01$ might however be a reminiscence of this slow helicoidal motion.

At higher $\mathrm{Ga}$, in the range $\mathrm{Ga}\in(225, 235)$, the perpendicular velocity fluctuations PSD presented in Fig.~\ref{fig:fft}(c) have a marked peak at the frequency $f^\ast = (0.055\pm0.010)$ with a broad base extending towards lower frequencies, down to the spectral resolution of 0.1. This behavior is similar to the one reported by Raaghav~\etal~\cite{breugem_new} for particles with density ratio $\Gamma \approx 3.9$ at $\mathrm{Ga}\sim 210$, where a peak at $f^\ast \simeq 0.05$ and a peak at $f^\ast \simeq 0.005$ were reported. This was interpreted as a probable superposition of Low-Frequency oblique oscillations and a slow helical rotation. This scenario is consistent with the combined analysis of angle, planarity and spectral content in the present study. Indeed Fig.~\ref{fig:angle}(a) shows that trajectories in the range $\mathrm{Ga}\in(225, 235)$ are oblique, while Fig.~\ref{fig:planarity_met} indicates coexistence of planar and non-planar (hence compatible with helical motion) trajectories in this range of $\mathrm{Ga}$. Intriguingly while both, Raaghav \etal's~and the present experiments seem to observe this co-existence of Low-Frequency Oblique and helicoidal trajectories for high density ratio particles, such a behavior has not been reported in numerical simulations by Zhou and Du\v{s}ek~\cite{zhoupaper}.  

At the largest $\mathrm{Ga}$ explored, Fig.~\ref{fig:fft}(d) presents the PSDs for the case $\mathrm{Ga}$ = 235. It does not present any dominant frequency, as it is expected for Chaotic dynamics. 

Overall, our study of oscillations for the high density ratio particles ($\Gamma\approx 7.9$), is in good agreement with numerical simulations apart from the range $\mathrm{Ga}\in(227,233)$ where Low-Frequency oscillations, possibly co-existing with non-planar helical motion, were observed but not reported in simulations. Reasonable agreement is also found with previous experiments by Raaghav \textit{et al.} at density ratio $\Gamma \approx 3.9$, although we do confirm the existence of the High-Frequency oscillating region for $\mathrm{Ga}\in(187,205)$, which they did not observe, but is predicted by the simulations by Zhou and Du\v{s}ek~\cite{zhoupaper}. We do not observe however the same regimes as in the study by Veldhuis and Biesheuvel~\cite{veldhuis} at $\Gamma \approx 2.5$; in particular in the Oblique Oscillating Regimes, they report a Low-Frequency behavior (at $f^\ast \approx 0.07$) rather than a High-Frequency one, as predicted by the simulations. It is likely that is due to the fact that the density ratio they considered is very close to the Low/High-Frequency transition, found to occur around $\Gamma \approx 2.3$ in the simulations.\\

Fig.~\ref{fig:fft_PandG} presents PSDs of velocity fluctuations for $\Gamma \approx 1.1$ and $\Gamma \approx 2.5$ Particles, at different values of the Galileo number corresponding to the following data sets: $l^\ast_\mathrm{max} = 23.3$ for $\Gamma \approx 2.5$ particles; and $l^\ast_\mathrm{max} = 11.6$ for $\Gamma \approx 1.1$ particles.
Both parallel and perpendicular components of velocity fluctuations have been analyzed. Each sub-figure presents the ensemble average of all the PSDs in the following $\mathrm{Ga}$ regimes:
$\Gamma \approx 2.5$ Particles in the L-F Oscillating Regime showed in Fig.~\ref{fig:fft}(a); and $\Gamma \approx 1.1$ Particles in the L-F Oscillating and Vertical Periodic Regimes, presented in Fig.~\ref{fig:fft_PandG}(b) and (c), respectively. 

The perpendicular velocity fluctuations PSDs presented in Fig.~\ref{fig:fft_PandG}\fcb{(b)} show that for $\mathrm{Ga} = 208$ oscillations have a broad frequency peak centred around a dominant frequency $f^{\ast} = (0.043\pm0.021)$. While the parallel velocity presents a peak at the same frequency but with 10 times less energy.
Note that the uncertainty is considerably higher here since the trajectories are shorter ($l^\ast_\mathrm{max}$ = 11.6 or 23.3). The dominant frequency confirms the Low-Frequency nature of the oscillations qualitatively identified in the velocity signal shown in Fig.~\ref{fig:vperp_highvslow}. 
This is also in agreement with the frequency predicted by numerical simulations by Zhou and Du\v{s}ek~\cite{zhoupaper} and observations from Veldhuis and Biesheuvel~\cite{veldhuis}.

On the other hand, Fig.~\ref{fig:fft_PandG} \fcb{(c)} presents the perpendicular and parallel velocity PSDs of $\Gamma \approx 1.1$ particles in the range $\mathrm{Ga} \in (269,~272)$. We observe a single broad frequency peak centered around $f^{\ast} = (0.085\pm 0.021)$, that overlaps with the Low-Frequency. As for the sub-figure (a), the parallel velocity presents a peak at the same frequency but with 10 times less energy. 
\fcb{This peak is at frequencies slightly lower than the frequency identified by Zhou and Du\v{s}ek~\cite{zhoupaper} for the vertical periodic regime (of the order of $f^*\approx 0.15$). Overall, given the small (although not strictly zero) angle previously reported for particles in this range of parameters, our observations are globally consistent with the existence of such a Vertical Periodic Regime.
It is worth noting that the experiments of Raaghav \textit{et al.} also measured non-strictly-zero angles of $0.3^{\circ}$ for similar range of parameters. Raaghav \textit{et al.} have measured a frequency of $f^{\ast} = 0.15$ very close to the numerical prediction by Zhou and Du\v{s}ek. Besides they have shown that in this region of the parameters space both Chaotic and Vertical Oscillating trajectories may co-exist. This may be a possible explanation for the broader than expected peak at lower frequency here measured; given the low spectral resolution of the present measurements (for this particular dataset) we might actually be seeing a combination of Chaotic (broad spectra) and Vertical Periodic trajectories (with, a priori, $f^{\ast} = 0.15$). } \\
Finally, Fig.~\ref{fig:fft_PandG}\fcb{(a)} presents the perpendicular and parallel velocity PSDs of $\Gamma \approx 2.5$ particles in the range $\mathrm{Ga} \in (190,~210)$. 
The perpendicular velocity fluctuations PSD presented in Fig.~\ref{fig:fft_PandG}(c) shows that oscillations have a broad frequency peak centred around a dominant frequency $f^{\ast} = (0.054\pm0.013)$. Additionally, note that, as the trajectories are longer than for $\Gamma\approx 1.1$ ($l^\ast_\mathrm{max} \in (33.3,~100)$), the uncertainty in this case is smaller (though still larger than for $\Gamma\approx 7.9$).
This spectral content is in agreement with the Low-Frequency Regime predicted in numerical simulations by Zhou and Du\v{s}ek~\cite{zhoupaper}, and what Veldhuis and Biesheuvel~\cite{veldhuis} have measured for particles in this area of the parameters space. A difference with the experiments of Veldhuis and Biesheuvel~\cite{veldhuis} is however seen as they have found harmonic contributions at around $f^\ast = 0.27$ \cite{veldhuis}.
\subsection{Settling Velocity \& Drag}\label{sec:drag_spheres}
In this last section we investigate the terminal settling velocity of the particles which results from the balance of the drag force and net gravity (i.e. gravity plus buoyancy). The measure of terminal velocity therefore allows to estimate the drag coefficient of the falling spheres and compare it to tabulated values for fixed spheres.

As previously discussed, the dimensional analysis of the problem of a sphere falling in a quiescent viscous fluid, yields two dimensionless control parameters: $\mathrm{Ga}-\Gamma$. When addressing the further question of the terminal vertical velocity $v_s$, an additional dimensionless parameter emerges: the terminal particle Reynolds number $\mathrm{Re_p} = v_sd_p/\nu$. It is important to note that $\mathrm{Re_p}$ is a response parameter of the problem which depends on the control parameters $\Gamma$ and $\mathrm{Ga}$ (we shall write then $\mathrm{Re_p}(\mathrm{Ga},\Gamma)$), therefore implying a possible impact of the path instabilities (which depend on both $\mathrm{Ga}$ and $\Gamma$) previously discussed on the terminal velocity of the spheres. Similarly, when it comes to address the question of the drag force experienced by the falling sphere, this introduces another dimensionless parameter, the drag coefficient $C_D$, which shall also be considered \textit{a priori} as a function of both $\mathrm{Ga}$ and $\Gamma$ (we shall write $C_D(\mathrm{Ga},\Gamma)$). This situation therefore contrasts with the case of the drag force of a fixed sphere in a prescribed mean stream, as in that situation, the density ratio is not a relevant parameter, and Reynolds number is then the unique control parameter of the problem. The drag coefficient solely depends in that case on the sphere Reynolds number $C_D(\mathrm{Re_p})$.\\
This then raises several points for the case of settling spheres: \\
(i) Are the usual correlations for the drag coefficient $C_D(\mathrm{Re_p})$ (not explicitly dependent on the density ratio $\Gamma$) still valid for the case of falling spheres (where $\mathrm{Re_p}$ and $\mathrm{C_D}$ may have explicit dependencies on both $\mathrm{Ga}$ and $\Gamma$)? Recall that explicit dependency on density ratio is known to be potentially major for light particles with $\Gamma \ll 1$ \citep{auguste_magnaudet_2018,karamanev}; \\
(ii) $\mathrm{Re_p}$ being a response parameter, usual correlations for the drag coefficient of fixed spheres $C_D(\mathrm{Re_p})$ are impractical as $\mathrm{Re_p}$ is not known beforehand: correlations directly implying the actual control parameters $(\mathrm{Ga},\Gamma)$ (eventually only $\mathrm{Ga}$ if explicit dependency on density ratio is found not to be important) would be more practical; \\
(iii) If density ratio is found to play a role, how important are the associated effects?\\ 
We address here these questions.
\subsubsection{New correlation relations between Galileo number and terminal particle Reynolds number / Drag coefficient}
Consider a settling particle within a given point of the parameters space $(\mathrm{Ga},\Gamma)$, with a terminal settling velocity $v_s(\mathrm{Ga},\Gamma)$. From the definition of the terminal particle Reynolds number $\mathrm{Re_p} = v_s d_p/\nu$ and of the Galileo number $\mathrm{Ga}=U_g d_p/\nu$, we can define the dimensionless particle terminal velocity $v_s^{\ast}$, which can be rewritten in terms of $\mathrm{Ga}$ and $\mathrm{Re_p}$ \cite{thesisfacu}:
\begin{equation}
v_s^{\ast} (\mathrm{Ga},\Gamma)= \frac{v_s}{U_g} = \frac{\mathrm{Re_p}(\mathrm{Ga},\Gamma)}{\mathrm{Ga}}.
\label{eq:vs_eq}
\end{equation}

Regarding drag, considering that in the terminal settling the drag force $F_D=\frac{1}{8}\rho_f C_D \pi d_p^2 v_s^2$ equals the gravity-buoyancy force $F_g = \frac{\pi}{6} (\rho_p-\rho_f)d_p^3 g = \frac{\pi}{6} \rho_f d_p^2 U_g^2$, from relation~\eqref{eq:vs_eq} the drag coefficient can be simply expressed as \cite{thesisfacu}: 
\begin{equation}
C_D(\mathrm{Ga},\Gamma) = \frac{4}{3}\bigg( \frac{\mathrm{Ga}}{\mathrm{Re_p}(\Gamma,\mathrm{Ga})}\bigg)^2.
\label{eq:drag_Ga_Re}
\end{equation}

Note that in this expression, the particle Reynolds number $Re_p(\mathrm{Ga},\Gamma)$ a response parameter of the problem, which is not known \textit{a priori} and needs to be measured. As further discussed below it can be analytically expressed only in the vanishing Galileo number limit, which corresponds to the steady vertical Stokes settling regime.

\begin{figure}[t!]
\centering
   \includegraphics[width=0.95\linewidth]{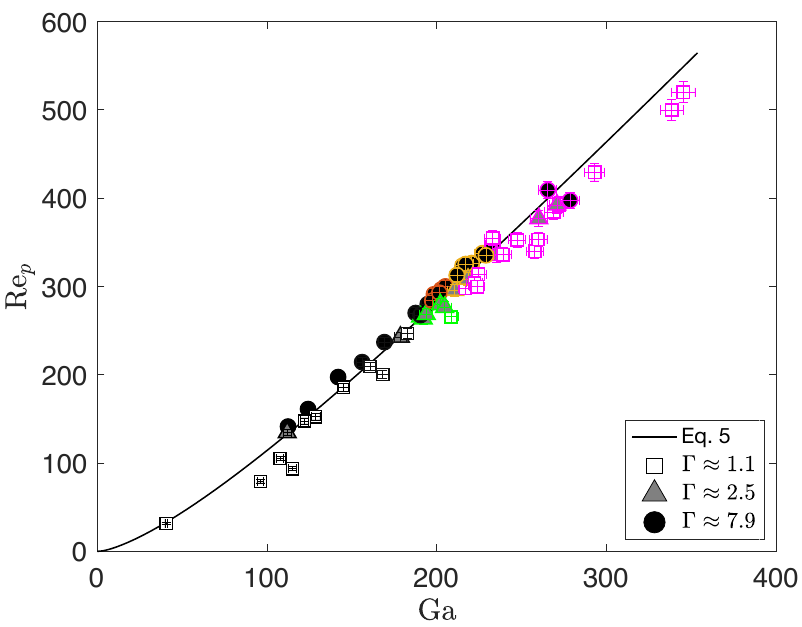}
    \caption{Galileo number versus particle Reynolds number alongside with the empirical correlation from Eq.~\ref{eq:Re_as_function_of_Ga}. The symbols represent the different density ratios (i.e. particle material): squares -- $\Gamma \approx 1.1$; triangles -- $\Gamma \approx 2.5$; circles -- $\Gamma \approx 7.9$. Whereas the edge colors represent the different trajectory regimes, as in Fig.~\ref{fig:parametersspace}: black -- Rectilinear \& Oblique; green -- Low-Freq.; orange -- High-Freq.; yellow -- Planar or Rotating; and magenta -- Chaotic \& Vertical Periodic.  }
   \label{fig:Ga_Re_notcomp}
\end{figure}

Fig.~\ref{fig:Ga_Re_notcomp} presents the measurements of $\mathrm{Re_p}$ versus Galileo number, for all particles (of all density ratios and for all the settling regimes) explored in the present study. The points appear to be relatively well packed on a main common trend, implying a minor direct dependency of $\mathrm{Re_p}$ on the density ratio $\Gamma$ (note that an implicit dependency on $\Gamma$ still exist via $\mathrm{Ga}=\sqrt{(\Gamma - 1)g d_p^3}/\nu$). Some scatter of the points is however visible, which may still reflect a possible explicit (minor) correction to the main trend due to the density ratio (this aspect will be further discussed in the next Subsection).

Before addressing such possible corrections, let first consider as a first approximation that $Re_p$ is independent on the density ratio and only explicitly dependent on $\mathrm{Ga}$. According to~\eqref{eq:drag_Ga_Re}, that implies then that the drag coefficient $C_D$ is itself also independent of the density ratio, and solely dependent on $\mathrm{Ga}$. Since $Re_p$ and $\mathrm{Ga}$ are then related, $C_D$ can be equivalently considered as $\mathrm{Ga}$-dependent or $\mathrm{Re_p}$-dependent. This is in agreement with previous studies by~\citep{horowitz,breugem_new} who measured the drag coefficient of falling spheres and did not observe, within the scatter of their measurements, a significant deviation compared to the fixed sphere case.

It can be noted that the empirical finding that neither $C_D$ nor $\mathrm{Re_p}$ explicitly depend on $\Gamma$, while they are univoquely related via $\mathrm{Ga}$, is trivial in the Stokes settling regime (in the limit of vanishing $\mathrm{Ga}$ and $\mathrm{Re_p}$). In this limit, analytical solutions of Stokes equations, lead indeed to $C_D(\mathrm{Re_p}) = 24/\mathrm{Re_p}$, that combined with Eq.~\ref{eq:drag_Ga_Re} yields $\mathrm{Re_p} = \frac{1}{18} \mathrm{Ga}^2$.

For non vanishing $\mathrm{Ga}$ and $\mathrm{Re_p}$, a univoque relation between $\mathrm{Re_p}$ and $\mathrm{Ga}$ supports then the idea that an explicit correlation $\mathrm{Re_p}(\mathrm{Ga})$ between these two parameters (via~\eqref{eq:Cd_as_function_of_Ga}) can be derived using classical correlations for $C_D(Re_p)$ for fixed spheres. We propose here to use the correlation by~\citep{brown}, which accurately fits the drag coefficient for spheres over a broad range of Reynolds number (up to $\mathrm{Re_p} \lesssim 2\times 10^5$): 

\begin{equation}
C_D(\mathrm{Re_p}) = \frac{24}{\mathrm{Re_p}}(1+ 0.150 \mathrm{Re_p}^{0.681}) + \frac{0.407}{1+\frac{8710}{\mathrm{Re_p}}} .
\label{eq:drag_brown}
\end{equation}

By including this expression of $C_D(Re_p)$ into~\eqref{eq:drag_Ga_Re}, we can indeed provide a direct correlation for the terminal particle Reynolds number (and hence for the particle terminal velocity) only depending on the actual control parameter of the problem which is the Galileo number:

\small
\begin{equation}
\mathrm{Re_p}^{\dagger}(\mathrm{Ga}) = \frac{\mathrm{Ga}^{2} (22.5 + {\mathrm{Ga}}^{1.364})}{0.0258{\mathrm{Ga}}^{2.6973} + 2.81{\mathrm{Ga}}^{2.0306} + 18{\mathrm{Ga}}^{1.364} + 405}.\\
\label{eq:Re_as_function_of_Ga}
\end{equation}
\normalsize

This expression is represented in Fig.~\ref{fig:Ga_Re_notcomp} by the solid line, and is found in very good agreement with the global trend measured for the settling particles in our experiments (what essentially confirms that the drag coefficient for fixed spheres reasonably applies to the case of falling spheres). Beyond this agreement, the above correlation is of great practical interest as it allows a direct determination of the settling velocity of a sphere from the sole \textit{a priori} knowledge of its Galileo number (which is a true control parameter, only requiring to know the particle-to-fluid density ratio, the sphere diameter, the acceleration of gravity and the ambient fluid's kinematic viscosity), without the need of using the traditional $\mathrm{C_D}(\mathrm{Re_p})$ correlation to solve (numerically) the non-linear equation~\eqref{eq:drag_Ga_Re}: $\mathrm{Re_p}^2\mathrm{C_D}(\mathrm{Re_p})=\frac{4}{3}Ga^2$.

Similarly, a direct correlation between the drag coefficient and the actual control parameter of problem ($\mathrm{Ga}$) (rather than the usual correlation $C_D(Re_p)$, which connects two response parameters) can be derived by re-introducing expression~\eqref{eq:Re_as_function_of_Ga} back into~\eqref{eq:drag_Ga_Re}:

\small
\begin{equation}
C_D^{\dagger}(\mathrm{Ga}) = \frac{4}{3}\bigg( \frac{0.0258 \mathrm{Ga}^{2.6973} + 2.81 \mathrm{Ga}^{2.0306} +18Ga^{1.364} + 405 }{\mathrm{Ga} (22.5 + \mathrm{Ga}^{1.364})} \bigg)^2.
\label{eq:Cd_as_function_of_Ga}
\end{equation}
\normalsize

\subsubsection{Density ratio effect}

The new correlations~\eqref{eq:Re_as_function_of_Ga} and \eqref{eq:Cd_as_function_of_Ga} we just proposed assume that both the terminal Reynolds number $Re_p$ and the drag coefficient $C_D$ only depend on $\mathrm{Ga}$ and do not depend explicitly on $\Gamma$. Based on Fig.~\ref{fig:Ga_Re_notcomp}, this seems a reasonable global assumption, though some scatter of the points in Fig.~\ref{fig:Ga_Re_notcomp} and small deviations (in particular for the less dense particles, $\Gamma \approx 1.1$ particles, represented as squares in the figure) with respect to relation~\eqref{eq:Re_as_function_of_Ga} cannot rule out a possible (minor) effect of density ratio. 

\begin{figure}[!htbp]
\centering
\includegraphics[width=0.95\linewidth]{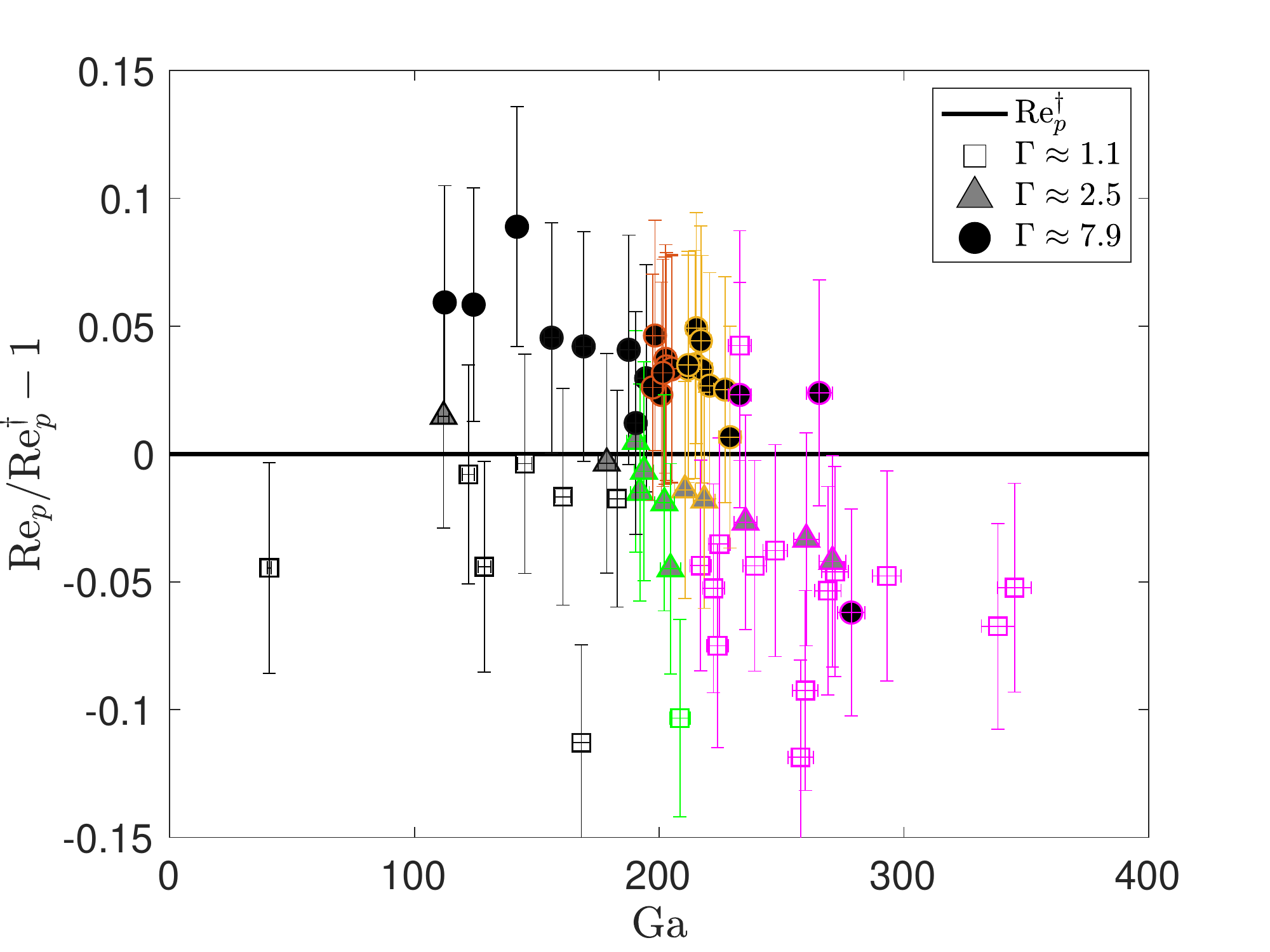}
\caption{Galileo number versus particle Reynolds number compensated by the empirical correlation from Eq.~\ref{eq:Re_as_function_of_Ga}. The symbols represent the different density ratios (i.e. particle material): squares -- $\Gamma \approx 1.1$; triangles -- $\Gamma \approx 2.5$; circles -- $\Gamma \approx 7.9$.  Whereas the edge colors represent the different trajectory regimes, as in Fig.~\ref{fig:parametersspace}: black -- Rectilinear \& Oblique; green -- Low-Freq.; orange -- High-Freq.; yellow -- Planar or Rotating; and magenta -- Chaotic \& Vertical Periodic. }
   \label{fig:Ga_Re_comp}
	\end{figure}

\begin{figure}[!htbp]
\centering
\includegraphics[width=0.95\linewidth]{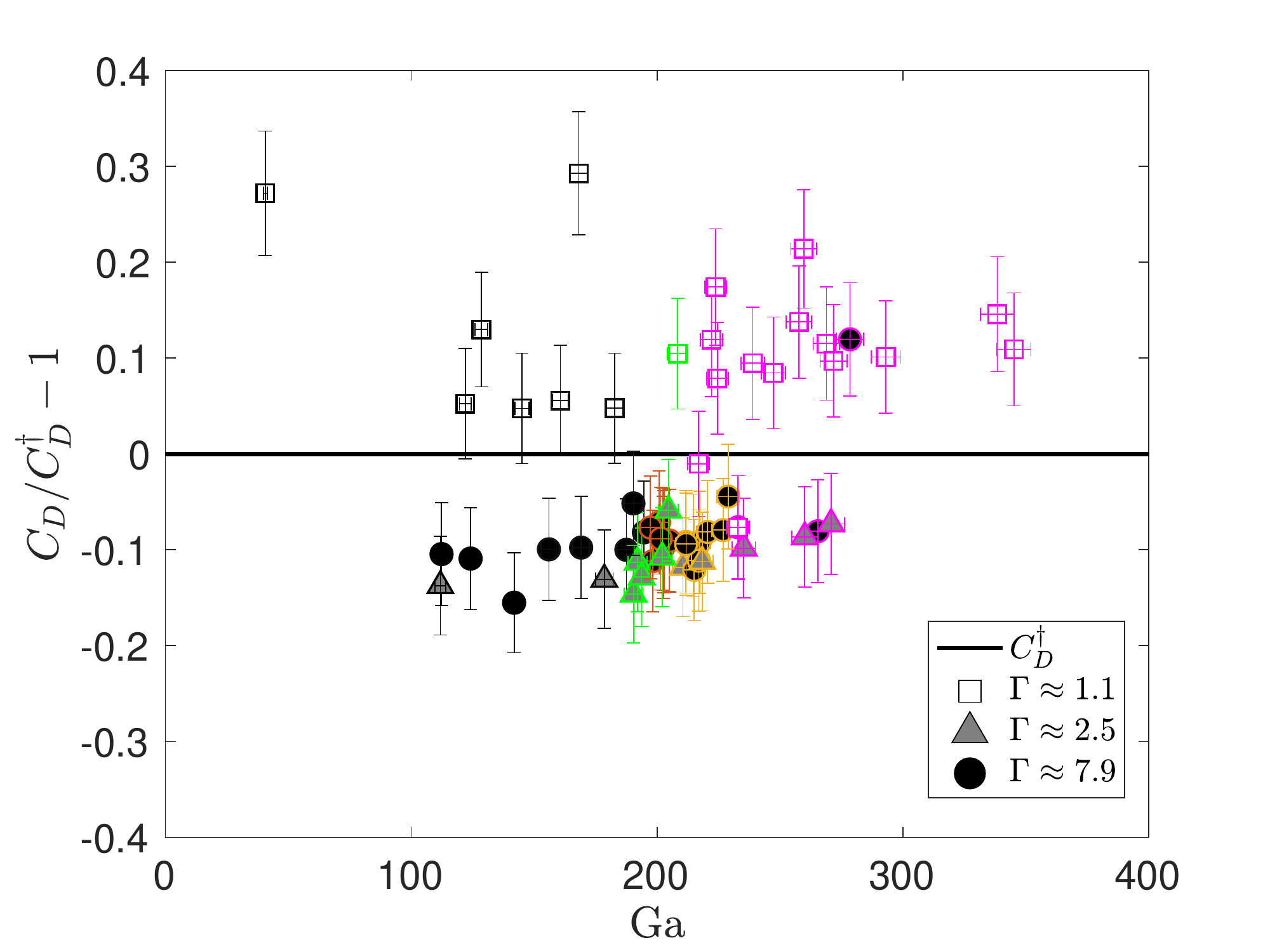}
\caption{Drag coefficient compensated by the empirical correlation from Eq.~\ref{eq:Cd_as_function_of_Ga} versus Galileo number. The symbols represent the different density ratios (i.e. particle material): squares -- $\Gamma \approx 1.1$; triangles -- $\Gamma \approx 2.5$; circles -- $\Gamma \approx 7.9$. Whereas the edge colors represent the different trajectory regimes, as in Fig.~\ref{fig:parametersspace}: black -- Rectilinear \& Oblique; green -- Low-Freq.; orange -- High-Freq.; yellow -- Planar or Rotating; and magenta -- Chaotic \& Vertical Periodic. }
\label{fig:Cd_spheres_comp}
\end{figure}

To better test possible deviations due to density ratio effects, we show in Fig.~\ref{fig:Ga_Re_comp} and \ref{fig:Cd_spheres_comp} the terminal Reynolds number and the drag coefficient compensated respectively by relations~\eqref{eq:Re_as_function_of_Ga} and~\eqref{eq:Cd_as_function_of_Ga} such that a value of zero would correspond to a perfect match (hence with no density effects). 

Fig.~\ref{fig:Ga_Re_comp} (for the compensated terminal Reynolds number) shows that although the measurements for all different datasets obtained in this work are indeed distributed around zero, they can deviate from this density-independent trend with a scatter of typically $\pm10\%$. More importantly it can be seen that (apart for two outliers out of the 68 independent measurements we carried) the scatter of the points present a systematic trend with the density ratio, where less dense particles (notably $\Gamma \approx 1.1$ particles and, to a less extent, $\Gamma \approx 2.5$ particles) are systematically below the correlation derived from fixed spheres, while heavy particles are systematically above. The density-independence approximation seems therefore to give a reasonable average trend to predict the terminal Reynolds number using relation~\eqref{eq:Re_as_function_of_Ga} though denser particles will have a positive bias (settling up to 10\% faster in the range of densities explored here) and lighter particles a negative bias (up to 13\% slower in the rage of densities explored here).

Similarly Fig.~\ref{fig:Cd_spheres_comp} shows that (apart for the same two outliers out of the 68 independent measurements we carried), a systematic effect of density ratio can be observed on the drag coefficient $\mathrm{C_D}$, where less dense particles (notably $\Gamma \approx 1.1$ particles) have a systematic positive bias (\textit{i.e.} their drag coefficient is larger, up to +15\% in the range of densities we explored) compared to the correlation derived from fixed spheres, while heavy particles have systematic negative bias (\textit{i.e.} their drag coefficient is lower, up to -15\% in the range of densities we explored) compared to the correlation derived from fixed spheres. The overall drag coefficient spread is 30\%. 


These results challenge the widespread idea that the drag coefficient (and eventually then its connection to the terminal settling velocity via relation~\eqref{eq:Cd_as_function_of_Ga}) of freely settling spheres (\textit{i.e.} with $\Gamma > 1$) do not explicitly depend on the density ratio $\Gamma$. 
\fcb{Previous studies are however not in contradiction with this claim~\cite{auguste_magnaudet_2018, horowitz, breugem_new, veldhuis}. Indeed, while these studies did not specifically focus on a quantitative estimation of possible fine deviations from the fixed sphere case, small systematic deviations can actually be observed in the reported data.
In particular,} we find a systematic explicit dependence on $\Gamma$, as $C_D$ and $\mathrm{Re_p}$ vary in 25\% to 30\% between the less dense ($\Gamma\approx 1.1$) and the denser particles ($\Gamma\approx 7.5$). It is worth to remark that the results from the denser particles ($\Gamma \approx 7.9$ particles) and the intermediate density ratio ones ($\Gamma \approx 2.5$ particles) are hardly distinguishable (in particular regarding the drag coefficient in Fig.~\ref{fig:Cd_spheres_comp}). This suggests that the $\Gamma$ dependency might be most relevant for $\Gamma$ values close to one, i.e. closer to the rising particle case where a clear dependency with $\Gamma$ was reported for the drag coefficient~\citep{karamanev, auguste_magnaudet_2018} and has been found to be systematically larger compared to the case of fixed spheres. Deviations for light particles with $\Gamma\lesssim 1$ remain small and comparable to the ones we report here for $\Gamma \approx 1.1$ particles with $\Gamma \gtrsim 1$, and become important for very light spheres with $\Gamma \ll 1$.
\section{Conclusions}\label{sec:conclusions}
We presented in this article an experimental study on the settling of single spheres in a quiescent flow, with a systematic characterization of settling regimes, settling terminal velocity and drag coefficient of spheres with density ratios up to $\Gamma \simeq 8$ (previous similar studies were limited to $\Gamma < 4$). The spheres dynamics is analyzed in the parameters space $\Gamma-\mathrm{Ga}$, with particle-to-fluid density ratios $\Gamma \in (1.1,7.9)$ and Galileo numbers $\mathrm{Ga}\in (100,340)$. 

Overall, our results on the settling regimes are in very good agreement with the numerical simulations by Zhou and Du\v{s}ek~\cite{zhoupaper} and in partial agreement with previous experiments by \VB~\cite{veldhuis} and Raaghav \textit{et al.}~\cite{breugem_new} over a narrower range of density ratios.

In particular, we confirm that for all situations, trajectories eventually become chaotic in the high Galileo number limit (typically for $\mathrm{Ga}>250$) although the details of the route to chaos depends on the density ratio of the particles. For the lowest density ratio, we observe all the regimes predicted by Zhou and Du\v{s}ek~\cite{zhoupaper} simulations. In particular we confirm the Low-Frequency nature of Oblique Oscillating Regime (for $\mathrm{Ga}\lesssim 200$ for $\Gamma=1.1$ and around $\mathrm{Ga}\approx 200$ for $\Gamma=2.5$) with a dominant dimensionless frequency $f^\ast \approx 0.06$. While this regime (predicted by Zhou and Du\v{s}ek~\cite{zhoupaper}) was reported by Raaghav \textit{et al.}~\cite{breugem_new}, it was not clearly observed in experiments by \VB. We also confirm that particles with density ratio close to unity (Plastic Particles with $\Gamma = 1.1$) exhibit a ``pocket'' of vertical periodic settling in the range $\mathrm{Ga}\in(250,300)$. This regime predicted in simulations by Zhou and Du\v{s}ek~\cite{zhoupaper} was also reported in experiments by Raaghav~\textit{et al.} although it was not observed by \VB.

For the densest particles we investigated (Metallic Particles with $\Gamma = 7.9$), which are also the densest reported for such experimental studies, we confirm the existence of a High-Frequency Oblique Oscillating Regime, around $\mathrm{Ga}\approx 200$ with $f^\ast \approx 0.18$. This regime was not observed in experiments Raaghav \textit{et al.}~\cite{breugem_new} at $\Gamma=3.9$ who only reported helical/rotating trajectories. We also observe such helical trajectories (around $\mathrm{Ga}\approx 220$), which we find to co-exist with the High-Frequency Oblique Oscillating Regime for $\mathrm{Ga}\lesssim 220$, in agreement with what Zhou and Du\v{s}ek~\cite{zhoupaper} identified as a mult-stable Planar-or-Rotating Regime, where both planar (oblique oscillating trajectories) and non-planar (helical trajectories) could be observed. We find however that the range of multi-stability is probably larger than what is reported in the numerical study by Zhou and Du\v{s}ek~\cite{zhoupaper}, as helicoids were randomly observed over almost the entire range of Galileo numbers \textit{a priori} corresponding to the High-Frequency Oblique Oscillating Regime. This may explain why the High-Frequency Oblique Oscillating Regime was not reported in~\cite{breugem_new}, who may have only (randomly) observed helical trajectories in this range. Concerning the helical trajectories, although the limited extent of the measurement volume in our experiment did not allow to fully characterize the helical properties, raw estimates of the radius (about 7 particle diameters) and the pitch (several hundreds particle diameters) of the portion of helicoids we observed are consistent with previous values reported in experiments by Raaghav \textit{et al.}~\cite{breugem_new} and simulations by Zhou and Du\v{s}ek~\cite{zhoupaper}.

Finally, our study of the spheres terminal settling velocity ($v_s$) and drag coefficient $C_D$ carries two important results. First, neglecting density ratio dependencies, we have proposed two new correlations directly relating the terminal Reynolds number $\mathrm{Re_p} = v_s d_p/\nu$ and the drag coefficient $C_D$ to the Galileo number $\mathrm{Ga}$. For the case of settling spheres, these relations are more handy to use compared to classical correlations between the $C_D$ and $\mathrm{Re_p}$ as, contrary to $\mathrm{Ga}$ which is a true control parameter of the problem, $\mathrm{Re_p}$ is a response parameter which cannot be determined beforehand. Secondly, we have shown that the usual approximation to neglect an explicit dependency on the density ratio $\Gamma$ (other than the implicit dependency through $\mathrm{Ga}$ of the terminal Reynolds number and drag coefficient) for settling spheres is not justified from the dimensional analysis and not fully supported by experimental findings. In particular, a trend was observed were the drag coefficient of the lightest particles was systematically larger than for the densest particles, with a difference up to about 30\% over the entire range of parameters we investigated. This indicates that, at least in the range of Galileo numbers explored here (with rich and complex settling regimes), while using the drag coefficient from usual correlations tabulated for fixed spheres (which can be considered as infinitely dense) at the corresponding Reynolds number may give the good order of magnitude of the terminal velocity, an accurate estimate would require to account for finite density ratio effects. Beyond the case of spheres settling in quiescent fluid addressed here, such corrections may also play a role in the context of modeling the drag force coupling of finite size inertial particles advected and settling in~turbulent~flows.

\section{Acknowledgements}
We acknowledge the technical expertise and the help of V. Dolique for the use of the Scanning Electron Microscope. This work was supported by the French research program IDEX-LYON of the University of Lyon in the framework of the French program “Programme Investissements d’Avenir” (Grant No. hlR-16-IDEX-0005).

\bibliographystyle{plain}

\bibliography{refs.bib}

\end{document}